\documentclass[authoryear,11pt]{elsarticle} 

\usepackage{amsmath}
\usepackage{amssymb}
\usepackage[labelfont=bf]{caption}
\usepackage{subcaption}
\usepackage[dvipsnames]{xcolor}
\usepackage{graphicx}
\usepackage{booktabs}
\usepackage{rotating}  
\usepackage{float}     
\usepackage[colorlinks=true,linkcolor=Blue, urlcolor  = Blue, citecolor=Blue]{hyperref}%
\usepackage{eurosym}
\usepackage{xurl}
\usepackage[a4paper,margin=2cm]{geometry}

\usepackage{setspace}
\usepackage{tikz}
\usetikzlibrary{decorations.pathreplacing}
\usepackage{pgfplots}
\pgfplotsset{compat=1.17}

\usepackage{amsthm}
\theoremstyle{plain}
\newtheorem{proposition}{Proposition}
\theoremstyle{definition}
\newtheorem{assumption}{Assumption}
\newtheorem{definition}{Definition}

\usepackage{algorithm}
\usepackage{algpseudocode}

\graphicspath{ {./figs/} }

\begin{document}

\begin{frontmatter}

\title{Tax reform as a constrained optimization problem: a piecewise-linear framework and software implementation\tnoteref{t1}}
\tnotetext[t1]{We would like to thank the following people for helpful conversations and comments: Luka Boeskens, Jasper van Dijk, Yifan Duan, Joaquim Gromicho, Dick den Hertog, Elias de Korte, Romee Lind, Matthew Salganik, Dick van der Sluijs, Ole Teutloff, Vinzenz Ziesemer. We would also like to thank colleagues at the Centre for Information Technology Policy at Princeton University for helpful comments and support in realizing this paper.}


\author[aff1,aff2,aff3]{Mark Verhagen}
\ead{mv8058@princeton.edu}

\author[aff4,aff5]{Menno Schellekens}
\ead{m.schellekens@uva.nl}

\author[aff6]{Michael Garstka}
\ead{michael@garstka.org}

\affiliation[aff1]{organization={Nuffield College, University of Oxford},
            city={Oxford},
            country={United Kingdom}}

\affiliation[aff2]{organization={Centre for Information Technology Policy, Princeton University},
            city={Princeton, NJ},
            country={United States}}

\affiliation[aff3]{organization={Leverhulme Centre for Demographic Science, University of Oxford},
            city={Oxford},
            country={United Kingdom}}

\affiliation[aff4]{organization={Amsterdam Business School, University of Amsterdam},
            city={Amsterdam},
            country={The Netherlands}}

\affiliation[aff5]{organization={Ministerie van Financiën},
            city={The Hague},
            country={The Netherlands}}

\affiliation[aff6]{organization={Independent researcher},
            country={United Kingdom}}

\begin{abstract}
In many countries, income tax codes have grown into a complex tangle of interacting brackets, benefits, and deductions. Despite widespread calls for systematic reform, successful attempts at reform are rare. Part of the problem is the difficulty of designing viable reform proposals. Politically viable reform must offer hard guarantees on income effects, marginal rates, and budgetary cost. Existing microsimulation tools can evaluate a reform proposal but cannot generate one by themselves. We develop a framework that casts tax reform as a constrained optimization problem. We show that any statutory tax code satisfying four mild assumptions reduces to a finite-dimensional piecewise-linear function for each taxpayer group, so reform becomes a linear or mixed-integer linear program whose decision variables are legislatable parameters: rates, bracket cutoffs, and lump-sum transfers. We are able to recover current tax systems and generate provably optimal reform candidates within the modeled space, or a certificate that no reform satisfying certain policy design constraints exists. Behavioral effects can also be incorporated, producing a nonconvex mixed-integer formulation. We demonstrate the framework through a near-complete reconstruction of the Dutch income tax code, generating reforms that smooth marginal-rate spikes, cap household income losses, and roughly halve the number of active rules through a lexicographic procedure. Developed in close collaboration with the Dutch Ministry of Finance, the methodology is currently in active use there. An open-source software implementation is available as \texttt{TaxSolver}.

\end{abstract}


\end{frontmatter}

\section{Introduction}
\label{sec:intro}

In developed countries income taxation is a core pillar of government revenue and a central lever on economic behavior and equity~\citep{oecd2023revenue, mirrlees2011tax, tummers2019public, slemrod2013tax}. A large theoretical literature, culminating in the Mirrlees review, prescribes how rates should be set to balance efficiency and redistribution~\citep{mirrlees2011tax, Saez2001_Elasticities, GruberSaez2002_ETI, HuangRios2016_OptimalTaxMix,Mirrlees1971_OptimalTax}. Realized tax codes, however, often diverge sharply from what this theory would prescribe~\citep{mirrlees2011tax, slemrod2018tax}. A telling example is the Netherlands, where marginal rates spike above 80\% for many low- and middle-income households as several income-dependent benefits and credits phase out at once, sharply blunting work incentives~\citep{cnossen2022tax}. Comparable spikes, generally at odds with basic optimal-taxation principles, appear across developed countries and are typically unintended by-products of past reforms~\citep{Saez2001_Elasticities, slemrod2018tax}.

This mismatch has two root causes. First, implementing optimal-taxation prescriptions is hard even in principle: a tax code typically contains a multitude of interacting rules, and changing one part can have unintended consequences elsewhere. Second, reform is constrained by political reality~\citep{castanheira2012political, ilzetzki2018tax}: existing rules must be respected, and proposals often have to satisfy bespoke constraints that lie outside standard optimal-taxation toolkits~\citep{bierbrauer2016efficiency, feldstein1976theory}. The result is that reform tends toward ad-hoc adjustment of individual rules rather than systematic redesign, so reforms frequently miss, or even work against, their stated objectives~\citep{feldstein1976theory, diamond2011case, slemrod2018tax}. This ``implementation gap'' between optimal-tax theory and realized systems is well documented~\citep{mirrlees2011tax, slemrod2018tax, heathcote2017optimal}, and work in inverse optimal taxation finds that realized schedules often imply inconsistent social-welfare weights~\citep{bourguignon2006microsimulation, lockwood2016positive}.

We propose a framework that casts real-world tax reform as a constrained optimization problem. Under mild assumptions, any statutory income tax code can be written as a single piecewise-linear function per group of taxpayers. This lets us treat tax design as choosing the code's own parameters (rates, cutoffs, and lump-sum transfers) subject to a rich set of constraints that align with the political reality of tax reform.

In our setup, the decision variables are the legislatable parameters of the code and the constraints are the political and distributional policy requirements of the reform, separating our approach from existing computational traditions. First, tax--benefit microsimulation models evaluate reforms the analyst supplies, whereas we generate them. Second, numerical optimal-taxation models optimize abstract smooth schedules under simplistic assumptions, whereas our object is the statutory code itself. Third, reinforcement-learning tax design returns a black-box policy, whereas we return either a tractable optimal reform or a certificate that the commitments conflict. Section~\ref{sec:related} further develops this positioning.

Concretely, this paper makes the following contributions:
\begin{enumerate}
\item \textbf{A modeling result.} Under four mild assumptions (additivity, scalar inputs, piecewise linearity, group heterogeneity), any statutory income tax code reduces to a single piecewise-linear function per tax group whose breakpoints are the union of all rule cutoffs (Proposition~1). The function's parameters (rates, cutoffs, lump sums) are exactly the quantities legislation specifies.
\item \textbf{Inverse recovery with identifiability conditions.} Recovering the status-quo parameters from observed liabilities is a feasibility problem of inverse-optimization type and we give the rank condition under which recovery is unique (Proposition~2).
\item \textbf{A reform-optimization framework.} A library of constraints (two-sided income guarantees, marginal-rate caps, budget bands, protected rules), objectives (revenue loss, weighted rule-count cardinality), and a dynamic-bracketing formulation that makes the support itself a decision variable, each classified as a linear program (LP), mixed-integer linear program (MILP), or nonconvex mixed-integer quadratically constrained program (MIQCP) on first appearance, with infeasibility certificates.
\item \textbf{A representative case study.} A reconstruction of the Dutch income tax code (15 rules, 13{,}500 weighted taxpayers representing 13.5 million): a family of certified reforms that cap marginal pressure (the rate at which a euro of additional income is taxed) at 55--80\%, guarantee no household falls below 95\% of current net income within a $\pm$1.5\% budget band, and roughly halve the number of active rules via a lexicographic two-stage procedure, with every formulation remaining tractable at national scale (instance sizes, solve times, and optimality gaps are reported in the Supplementary Material).
\item \textbf{An open-source software implementation.} \texttt{TaxSolver}, a solver-agnostic Python package with a full reproducibility package.
\end{enumerate}

The remainder of this paper proceeds as follows. Section~\ref{sec:related} positions our contribution in the tax-policy and operations-research literatures. Section~\ref{sec:model} develops the piecewise-linear representation of statutory tax codes. Section~\ref{sec:reform} formulates tax reform as a constrained optimization problem over that representation and illustrates it on simple examples. Section~\ref{sec:cases} demonstrates the framework at national scale on the Dutch income tax code. Section~\ref{sec:practice} describes the open-source implementation and its use in practice, and Section~\ref{sec:discussion} concludes the paper.

\section{Related work}
\label{sec:related}

Our work draws on, and contributes to, two literatures: the economics of tax policy, which tells us what tax schedules \emph{should} look like in theory, and the operations-research methodology that provides the tools to optimize the schedule a statute actually encodes. We position the paper against each in turn.

\subsection{Tax policy background}
\label{sec:related_tax}

A large body of work in public economics studies what optimal tax rates ought to be, from the optimal-income-tax tradition initiated by \citet{Mirrlees1971_OptimalTax} to the sufficient-statistics approach built on elasticities of taxable income~\citep{Saez2001_Elasticities, piketty2013optimal}. Far less attention has gone to \emph{implementing} such rates in statutory tax codes. This has led to an ``implementation gap'' between theory and practice that can be observed in the UK~\citep{mirrlees2011tax}, the US~\citep{slemrod2018tax}, and many other developed countries~\citep{mankiw2009optimal,mirrlees2011tax}. The literature on inverse optimal taxation further shows that social-welfare outcomes implicit in realized schedules are frequently inconsistent with the stated goals of tax codes~\citep{bourguignon2006microsimulation, lockwood2016positive}. 

This is perhaps not surprising, as the standard computational tools for this implementation step are designed for a different purpose. Tax--benefit microsimulation models such as EUROMOD and TAXSIM~\citep{sutherland2013euromod, feenberg1993introduction} trace how a \emph{given} reform propagates through the tax code and support ``what-if''-style analyses, and newer rules-as-code platforms such as OpenFisca~\citep{openfisca} and PolicyEngine~\citep{woodruff2023policyengine} that encode statutes directly as executable code have the same evaluative purpose. These engines do not, by themselves, generate reforms, nor do they encode the bespoke political constraints that shape real reform processes. Our framework is complementary to them: it takes the same statutory code and the same microdata and \emph{searches} the space of implementable codes for one that meets stated objectives.

Reform generation has nonetheless been attempted from several directions. Numerical optimal-taxation and quantitative-macro studies optimize abstract, smooth schedules within stylized behavioral models~\citep{tuomala1990optimal, golosov2006new, conesa2006optimal}, and a neighboring strand restricts attention to small parametric piecewise-linear schedules~\citep{sheshinski1972optimal, slemrod1994optimal, apps2014optimal}. Closest in spirit to our work is a computational tradition that embeds an \emph{estimated} microsimulation model of household labour supply and performs a numerical search over parametric tax-transfer rules. \citet{aaberge2013using} identify optimal schedules for Norway, \citet{blundell2012employment} for low-income families in the United Kingdom, and \citet{colombino2022combining} extend the approach to polynomial rules across six European countries. \citet{kasy2018optimal} brings machine-learning estimation to the same design problem, and \citet{zheng2022ai} learn tax schedules from simulated economies by deep reinforcement learning. These approaches emphasize economic models of behavioral response but search over low-dimensional families of stylized rules rather than over the statutory code itself, and constraints beyond revenue neutrality are rare. Furthermore, search within this field is heuristic or learned, and so certifies neither optimality nor infeasibility.

Our framework differs on exactly these dimensions. Its decision variables are the legislatable parameters of the entire statutory code: in our Dutch application, fifteen interacting rules on multiple inputs rather than a single stylized schedule. Political, administrative, and distributional commitments enter as hard constraints rather than as terms in a welfare objective~\citep{bierbrauer2016efficiency}. Because the resulting programs are linear or mixed-integer linear, the output is either a reform that is certifiably optimal within the modeled space or a certificate that the stated commitments conflict. We also provide the option to encode behavioral responses at the cost of increased computational complexity, which we illustrate in Section~\ref{sec: case_3_behavioral}.

\subsection{OR foundations and related applications}
\label{sec:related_or}

Methodologically, our framework relies on several well-developed strands of operations research, each underwriting a specific part of the model.

\paragraph{Piecewise-linear functions and mathematical programming} 
Our framework leads to problem formulations that can be solved as linear programs or mixed integer programs. Linear programs have been studied since the 1930s by \citet{kantorovich1960mathematical} and~\citet{dantzig1951maximization} and fast polynomial time algorithms were developed by~\citet{karmarkar1984new}. While mixed-integer linear optimization is NP-hard (its decision version is NP-complete), steady algorithmic progress, beginning with the branch-and-bound method of \citet{land1960automatic}, has been made to solve problems with thousands of variables in a reasonable amount of time. A good overview is provided by~\citet{junger200950}.

Piecewise-linear functions feature heavily in applications that lead to optimization problems. Problems involving convex shaped piecewise-linear functions can be efficiently optimized. Optimizing over general piecewise-linear functions requires mixed integer programming and a modeling framework is presented by~\citet{vielma2010mixed} and~\citet{vielma2015mixed}. \citet{magnani2009convex} study the (nonconvex) problem of fitting a convex piecewise-linear function to data and give a heuristic that alternates data partitioning with least-squares fits. \citet{toriello2012fitting} show that fitting a general continuous piecewise-linear function to a finite set of data points can be solved using mixed-binary optimization.

\paragraph{Inverse optimization} Recovering the parameters of the current tax code from observed tax liabilities (Proposition~\ref{prop:recovery}) is an inverse problem: given the taxpayers' characteristics and taxes actually paid, infer the model that produced them. \citet{ahuja2001inverse} established that when the original problem is a linear program then the inverse problem is itself a linear program, which is why our recovery step is a linear feasibility problem. In our case we assume the availability of perfect reproducible tax data. If the tax data or the taxpayers' characteristics were noisy, then a perfect-fit model would likely not exist. Following the survey by \citet{chan2025inverse} there are several techniques for best-fit inverse problems to identify the tax system that best explains the noisy tax data.

\paragraph{Infeasibility diagnosis}
When a set of reform commitments cannot be met jointly, our framework returns a proof of infeasibility rather than a solution. Such infeasibility certificates are meaningful in and of themselves for policymakers, and isolating irreducible infeasible subsystems that name the conflicting commitments is a mature topic~\citep{chinneck1991locating, chinneck2008feasibility}. We illustrate their use in Section~\ref{sec: case_4_dutch} to further diagnose conflicting commitments.

\paragraph{Cardinality objectives via mixed-integer optimization} The complexity objective of Section~\ref{sec:extensions}, minimizing a weighted count of active tax rules, can be implemented with binary indicator variables and a ``Big M'' term. \citet{natarajan1995sparse} showed that this type of cardinality term makes an otherwise convex problem NP-hard. Efficient handling of such objectives in practice are the subject of a substantial mixed-integer-optimization literature; \citet{bertsimas2016best} show that its canonical instance, best-subset selection, is solvable at scale with modern MILP solvers. Tax code simplicity is often a key objective of real-world reforms~\citep{chetty2009salience, mirrlees2011tax}.

\paragraph{Lexicographic optimization with slack} A tax reform usually pursues several goals at once, e.g. minimal revenue loss \emph{and} minimal system complexity, that cannot easily be formulated as a single scalar objective. \citet{ehrgott2005multicriteria} discusses different approaches to solve multi-objective decision problems. We use a lexicographic approach with slack at each objective level. The two-stage Dutch tax system procedure of Section~\ref{sec:cases} first minimizes revenue loss and then, subject to a bounded revenue slack, minimizes tax rule complexity, so that a controlled amount of revenue can be spent to buy simplicity. The result is an $\varepsilon$-constraint-style traversal of the tax revenue--complexity frontier.

\paragraph{OR in government and policy analytics} Our application sits within a long tradition of operations research applied to public-sector and government decision-making. 
Several works use linear programming or mixed-integer programming to allocate or distribute public goods or facilities: \citet{toregas1971location} to determine the best location of emergency service facilities, \citet{cardoso2015integrated} to design long-term care networks, \citet{fishbone1981markal} to design energy systems, and \citet{barbarosoglu2004two} to plan effective disaster response.

Some recent work optimizes statutory and policy instruments directly: \citet{chintapalli2022implications} analyze a government minimum-support-price scheme under heterogeneous (myopic and strategic) behavioral responses of farmers which is structurally similar to our extension to incorporate behavioral response of taxpayers discussed in Section~\ref{sec: case_3_behavioral}. 

What distinguishes the present paper from tax-related literature is its decision object. A considerable body of work optimizes decisions taken \emph{under} a fixed tax system: post-tax portfolio selection moves statutory tax rules into the investor's allocation problem~\citep{osorio2004posttax}; global supply-chain network design chooses facility locations and transfer prices to maximize after-tax profit subject to national tax rules~\citep{vidal2001global}; and a range of corporate-tax-planning models schedule decisions such as loss carry-forwards to minimize tax owed. In each case the tax code is data: a fixed constraint on some other agent's optimization. We invert this relationship and make the statutory code itself the decision variable, optimized subject to the guarantees a reform must honor. We are not aware of prior OR work that takes the statutory income-tax code itself as the object of decision.

\section{A piecewise-linear model of statutory tax codes}
\label{sec:model}

We start this section with definitions that help us reason about a tax system and formulate reform as a constrained optimization problem. We define an income `tax code' as a set of `tax rules'. Each tax rule maps a `tax-relevant input' (input, hereafter) to an `absolute tax pressure' (the tax liability) and a `marginal tax pressure' (the effective marginal tax rate). Formally, an individual tax rule $r$ can be described as:

\begin{equation}
y_{i,r} = f_r(x_i;\, \boldsymbol{\pi}_r(c_i)),
\end{equation}
where $y_{i,r} \in \mathbb{R}$ is the absolute tax pressure for taxpayer $i$, $x_i \in \mathbb{R}$ is the input, and $\boldsymbol{\pi}_r(c_i)$ are parameters that may depend on taxpayer characteristics $c_i \in \mathbb{R}^P$.

An example tax rule could be an income tax bracket, with yearly income before tax as input. The absolute tax pressure is the amount of taxes payable or receivable due to this specific tax rule. The marginal tax pressure is the amount by which absolute tax pressure changes when the input increases by one unit. Finally, each taxpayer has `taxpayer characteristics' that impact what tax rules are applicable to them. These definitions are listed in Table \ref{table: tax_system_definitions}.

\begin{table}[!h]
\footnotesize
  \centering
  \renewcommand{\arraystretch}{1.3}
  \begin{tabular}{p{4cm} p{8cm}}
    \toprule
    \textbf{Term} & \textbf{Definition} \\
    \midrule
    Tax code &
    Complete set of tax rules applicable to taxpayers. \\

    \addlinespace
    Taxpayer &
    An individual who is charged taxes within a country’s income tax code. \\

    \addlinespace
    Input &
    A variable that is taxed (positively or negatively) within the tax code, such as income or capital gains. \\

    \addlinespace
    Tax rule &
    A formula that takes an input and returns an absolute amount of taxes payable or receivable. \\

    \addlinespace
    Absolute tax pressure &
    The absolute amount of taxes payable (the tax liability). \\

    \addlinespace
    Marginal tax pressure (with respect to an input) &
    The increase or decrease in absolute tax pressure when the input increases by one unit (the effective marginal tax rate). \\

    \addlinespace
    Taxpayer characteristics &
    The characteristics of a taxpayer that affect which types of tax rules are applicable to them. \\
    
    \bottomrule
  \end{tabular}
  \caption{Set of definitions to describe a tax system}
  \label{table: tax_system_definitions}
\end{table}

Individual tax rules are often simple. What makes tax codes complex is the number of tax rules in operation and how their shape and eligibility is governed by taxpayer characteristics. In fact, most tax systems consist of tax rules for which the following statements hold:

\begin{assumption}[Additivity]\label{ass:additivity}
The absolute and marginal tax pressures of all individual tax rules applicable to a taxpayer sum to that taxpayer's total absolute and marginal tax pressure.
\end{assumption}

\begin{assumption}[Scalar input]\label{ass:scalar}
Each tax rule levies taxes on a single input.
\end{assumption}

\begin{assumption}[Piecewise linearity]\label{ass:pwl}
For each tax rule, marginal pressure is constant on intervals of its input; equivalently, absolute pressure is a continuous, piecewise-linear function of that input.
\end{assumption}

\begin{assumption}[Group heterogeneity]\label{ass:hetero}
The eligibility and parametrization of a tax rule may depend on taxpayer characteristics, which partition the population into a finite set of tax groups.
\end{assumption}

We now verify that Assumptions~\ref{ass:additivity}--\ref{ass:hetero} hold for three archetype tax rules that cover the vast majority of tax rules in operation.

\subsection{Tax rules and the total tax code}

\paragraph{Tax brackets} Tax brackets are ubiquitous throughout most tax codes. A bracket rule can be described as:

\begin{equation}
f(x_i) = \sum_{b=1}^{B} \alpha_b\, g_b(x_i;\Phi), \qquad f : \mathbb{R} \to \mathbb{R},
\label{eq: tax_brackets}
\end{equation}
where $\alpha_b  \in \mathbb{R}$ is the marginal rate levied in bracket $b$ and $g_b(\cdot;\Phi)$ is the \emph{bracketing function} that returns the portion of the input falling in bracket $b$, defined as follows.

\begin{definition}[Bracketing function]\label{def:bracketing}
Let $\Phi = \{\phi_0, \phi_1, \dots, \phi_B\}$ be an ordered support with $0 = \phi_0 < \phi_1 < \cdots < \phi_B$ (with the convention $\phi_B = \infty$ for an open top bracket). For $b = 1, \dots, B$, the bracketing function is
\begin{equation*}
g_b(x;\Phi) = \max\bigl\{0,\; \min(x,\phi_b) - \phi_{b-1}\bigr\},
\end{equation*}
the amount of $x$ falling in the interval $(\phi_{b-1}, \phi_b]$. Each $g_b(\cdot;\Phi)$ is continuous and piecewise linear. We take $\phi_0 = 0$, so inputs are non-negative; then $\sum_{b=1}^{B} g_b(x;\Phi) = x$ for every $x \ge 0$ (and hence for every finite $x$ when $\phi_B = \infty$).
\end{definition}

We show two types of tax brackets in Figure \ref{fig: brackets}, both taking income before tax as input. The example on the left illustrates progressive brackets, where the marginal rate is equal to 10\% for incomes up until \euro25,000 and then increases on subsequent intervals. On the right, a flat income tax rate is shown with constant marginal pressure. Both tax rules satisfy Assumptions~\ref{ass:scalar} and~\ref{ass:pwl} (scalar input and piecewise linearity). As most income tax brackets are universal, the rates and cutoff points do not depend on taxpayer characteristics.

\begin{figure}[!t]
\centering
\definecolor{colorcurrent}{RGB}{0,114,178}   
\definecolor{colorreform}{RGB}{213,94,0}    
\definecolor{colorzzp}{RGB}{0,158,115}      
\begin{minipage}{0.47\textwidth}
  \centering
  \begin{tikzpicture}
    \begin{axis}[
        width=0.95\linewidth,
        height=0.8\linewidth,
        font=\small,
        xlabel={Income before tax (\euro)},
        xlabel style={font=\small},
        ylabel={Taxes paid (\euro)},
        ylabel style={font=\small},
        xmin=0, xmax=120000,
        ymin=0, ymax=75000,
        xtick={0,25000,50000,75000,100000},
        xticklabels={0,\euro25K,\euro50K,\euro75K,\euro100K},
        ytick={0,25000,50000,75000,100000},
        yticklabels={0,\euro25K,\euro50K,\euro75K,\euro100K},
        ticklabel style={font=\tiny,/pgf/number format/fixed},
        axis lines=left,
        grid=major,
        grid style={dashed, gray!60},
        enlargelimits=false,
        scaled ticks=false
      ]
      \addplot[
        thick,
        domain=0:120000,
        samples=200,
        color=colorcurrent
      ]
      { (x <= 25000) * (0.1 * x)
      + (x > 25000 && x <= 50000) * (2500 + 0.2 * (x - 25000))
      + (x > 50000 && x <= 75000) * (7500 + 0.3 * (x - 50000))
      + (x > 75000 && x <= 100000) * (15000 + 0.4 * (x - 75000))
      + (x > 100000) * (25000 + 0.5 * (x - 100000))
      };
    \end{axis}
  \end{tikzpicture}
\end{minipage}%
\hfill
\begin{minipage}{0.47\textwidth}
  \centering
  \begin{tikzpicture}
    \begin{axis}[
        width=0.95\linewidth,
        height=0.8\linewidth,
        font=\small,
        xlabel={Income before tax (\euro)},
        xlabel style={font=\small},
        xmin=0, xmax=120000,
        ymin=0, ymax=75000,
        xtick={0,25000,50000,75000,100000},
        xticklabels={0,\euro25K,\euro50K,\euro75K,\euro100K},
        ytick={0,25000,50000,75000,100000},
        yticklabels={},
        ticklabel style={font=\tiny,/pgf/number format/fixed},
        axis lines=left,
        grid=major,
        grid style={dashed, gray!60},
        enlargelimits=false,
        scaled ticks=false
      ]
      \addplot[
        thick,
        domain=0:120000,
        samples=100,
        color=colorcurrent
      ]
      {0.3 * x};
    \end{axis}
  \end{tikzpicture}
\end{minipage}
\caption{Illustration of two types of tax brackets. On the left, a 5‐bracket progressive tax (10\%, 20\%, 30\%, 40\%, 50\%) at intervals of \euro25,000; on the right, a single flat 30\% rate.}
\label{fig: brackets}
\end{figure}

\paragraph{Benefits} Benefits are another common tax rule that usually reflect a reduction in taxes paid. Benefits are often tailored to specific population groups for which eligibility and amounts can vary. The amount of reduction in taxes might also depend on an input like one's income before tax. In such cases, the benefit amount is typically `nullified' over an interval of the input.

The above means that the benefit tax rule can be viewed as a generalization of (\ref{eq: tax_brackets}) that include some lump sum transfer and eligibility criteria:

\begin{equation}
f(x_i) =
\begin{cases}
Z_1 + \sum_b \alpha_{b, 1} g_b(x_i; \Phi_{r, 1}), & \text{if } c_i \in \mathcal{C}_1 \\
Z_2 + \sum_b \alpha_{b, 2} g_b(x_i; \Phi_{r, 2}), & \text{if } c_i \in \mathcal{C}_2 \\
\vdots \\
Z_L + \sum_b \alpha_{b, L} g_b(x_i; \Phi_{r, L}), & \text{if } c_i \in \mathcal{C}_L,\\
0, &\text{otherwise}.
\end{cases}
\label{eq: benefits}
\end{equation}
where, for eligibility groups $\ell = 1, \dots, L$, $Z_\ell \in \mathbb{R}$ is a lump sum transfer, $\alpha_{b,\ell}  \in \mathbb{R}$ are the rates, $\Phi_{r,\ell}$ are the cutoff points, and $\mathcal{C}_\ell \subseteq \mathbb{R}^P$ is the set of characteristics that determine whether taxpayer $i$ belongs to group $\ell$.

\begin{figure}[!t]
  \centering
  \definecolor{colorcurrent}{RGB}{0,114,178}   
  \definecolor{colorreform}{RGB}{213,94,0}    
  \definecolor{colorzzp}{RGB}{0,158,115}      
  \makebox[\textwidth][c]{%
  \begin{minipage}{0.32\textwidth}
    \centering
    \begin{tikzpicture}
      \begin{axis}[
        width=\linewidth,height=0.90\linewidth,font=\small,
        xlabel={Income before tax (\euro)},ylabel={Taxes paid (\euro)},
        xmin=0,xmax=120000,ymin=-15000,ymax=15000,
        xtick={0,50000,100000},
        xticklabels={0,\euro50K,\euro100K},
        ytick={-10000,0,10000},
        yticklabels={-\euro10K,0,\euro10K},
        ticklabel style={font=\tiny,/pgf/number format/fixed},
        scaled ticks=false,axis lines=left,
        grid=major,grid style={dotted,gray!40},clip=false,
        legend style={
          at={(rel axis cs:0.97,0.97)},anchor=north east,
          draw=none,font=\tiny,
          legend columns=1
        }
      ]
        \addplot[thick,colorcurrent] coordinates{(0,-12000)(120000,-12000)};
        \addlegendentry{Group 1}
        \addplot[thick,colorreform] coordinates{(0,-10000)(120000,-10000)};
        \addlegendentry{Group 2}
      \end{axis}
    \end{tikzpicture}
  \end{minipage}\hspace{0.02\textwidth}%
  \begin{minipage}{0.32\textwidth}
    \centering
    \begin{tikzpicture}
      \begin{axis}[
        width=\linewidth,height=0.90\linewidth,font=\small,
        xlabel={\# children},
        xmin=0,xmax=8,ymin=-17000,ymax=15000,
        xtick={0,2,4,6,8},
        ytick={-10000,0,10000},
        yticklabels={},
        ticklabel style={font=\tiny,/pgf/number format/fixed},
        scaled ticks=false,axis lines=left,
        grid=major,grid style={dotted,gray!40},clip=false,
      ]
        \addplot[thick,colorcurrent,const plot mark left,mark=none] coordinates{
          (0,0)(1,-2000)(2,-4000)(3,-6000)(4,-8000)
          (5,-10000)(6,-12000)(7,-14000)(8,-16000)};
      \end{axis}
    \end{tikzpicture}
  \end{minipage}\hspace{0.02\textwidth}%
  \begin{minipage}{0.32\textwidth}
    \centering
    \begin{tikzpicture}
      \begin{axis}[
        width=\linewidth,height=0.90\linewidth,font=\small,
        xlabel={Income before tax (\euro)},
        xmin=0,xmax=120000,ymin=-15000,ymax=15000,
        xtick={0,50000,100000},
        xticklabels={0,\euro50K,\euro100K},
        ytick={-10000,0,10000},
        yticklabels={},
        ticklabel style={font=\tiny,/pgf/number format/fixed},
        scaled ticks=false,axis lines=left,
        grid=major,grid style={dotted,gray!40},clip=false,
        legend style={
          at={(rel axis cs:0.97,0.97)},anchor=north east,
          draw=none,font=\tiny,legend columns=1
        }
      ]
        \addplot[thick,colorcurrent] coordinates{
          (0,-12000)(50000,-12000)(100000,0)(120000,0)};
        \addlegendentry{Group 1}

        \addplot[thick,colorzzp] coordinates{
          (0,-8000)(50000,-8000)(75000,0)(120000,0)};
        \addlegendentry{Group 3}
      \end{axis}
    \end{tikzpicture}
  \end{minipage}}
  \caption{Example illustration of three types of benefits. On the left a group-dependent, universal benefit. In the middle, a childcare benefit that increases with the number of children. On the right, an income-dependent, group-dependent benefit.}
  \label{fig: benefits}
\end{figure}

Three examples of benefits are depicted in Figure \ref{fig: benefits}. The left-most inset shows two benefits that differ in the lump sum starting amount between two groups. Neither is dependent on an input and so there are no subsequent brackets. The middle inset shows a benefit that is again universal. The benefit increases linearly with the amount of dependent children one has, which can be seen as one large bracket with children as input. Finally, the right-most inset illustrates a benefit that starts at a lump sum value but takes income before taxes as input and is nullified over an interval. The height of the benefit and the interval on which it is nullified differs by group. It is straightforward to see that the assumptions listed earlier are again satisfied.

\paragraph{Deductibles} The third archetype tax rule is the deductible. A deductible represents a reduction in the amount of taxes payable and is usually implemented in the context of existing tax brackets. Two types are typically encountered. One type reduces the extent to which an input is taxed, for example reducing the amount of taxable income by some amount $D$. This is illustrated in Figure \ref{fig: deductibles} on the left in the context of a tax bracket rule. This type of deductible effectively shifts all cutoff points in (\ref{eq: tax_brackets}) up by $D$ and adds a ‘zero’ bracket that levies no taxes on the interval $[0, D]$. We call these `input-reducing deductibles'.

\begin{figure}[!t]
  \centering
  \definecolor{colorcurrent}{RGB}{0,114,178}   
  \definecolor{colorreform}{RGB}{213,94,0}    
  \definecolor{colorzzp}{RGB}{0,158,115}      
  \begin{minipage}{0.48\textwidth}
    \centering
    \begin{tikzpicture}
      \begin{axis}[
        width=\linewidth,
        height=0.8\linewidth,
        font=\small,
        xlabel={Income before tax (\euro)},
        ylabel={Taxes paid (\euro)},
        xmin=0, xmax=120000,
        ymin=0, ymax=75000,
        xtick={0,25000,50000,75000,100000},
        xticklabels={0,\euro25K,\euro50K,\euro75K,\euro100K},
        ytick={0,25000,50000,75000,100000},
        yticklabels={0,\euro25K,\euro50K,\euro75K,\euro100K},
        ticklabel style={font=\tiny,/pgf/number format/fixed},
        scaled x ticks=false, scaled y ticks=false,
        axis lines=left,
        grid=major,
        grid style={dashed,gray!60},
        clip=false,
        legend pos=north east,
        legend style={font=\tiny,draw=none}
      ]

        \addplot[thick,colorcurrent,domain=0:120000,samples=200] {
          (x<=25000)*(0.1*x)
        + (x>25000 && x<=50000)*(2500 + 0.2*(x-25000))
        + (x>50000 && x<=75000)*(7500 + 0.3*(x-50000))
        + (x>75000 && x<=100000)*(15000 + 0.4*(x-75000))
        + (x>100000)*(25000 + 0.5*(x-100000))
        };
        \addlegendentry{Original}
        
        \addplot[thick,colorreform,domain=0:120000,samples=200] {
          (x<=20000)*0
        + (x>20000 && x<=45000)*(0.1*(x-20000))
        + (x>45000 && x<=70000)* (2500 + 0.2*(x-45000))
        + (x>70000 && x<=95000)* (7500 + 0.3*(x-70000))
        + (x>95000 && x<=120000)*(15000 + 0.4*(x-95000))
        + (x>120000)* (25000 + 0.5*(x-120000))
        };
        \addlegendentry{Input-reduced}
      \end{axis}
    \end{tikzpicture}
  \end{minipage}\hfill
  \begin{minipage}{0.48\textwidth}
    \centering
    \begin{tikzpicture}
      \begin{axis}[
        width=\linewidth,
        height=0.8\linewidth,
        font=\small,
        xlabel={Income before tax (\euro)},
        xmin=0, xmax=120000, ymin=0, ymax=75000,
        xtick={0,25000,50000,75000,100000},
        xticklabels={0,\euro25K,\euro50K,\euro75K,\euro100K},
        ytick={0,25000,50000,75000,100000},
        yticklabels={0,\euro25K,\euro50K,\euro75K,\euro100K},
        ticklabel style={font=\tiny,/pgf/number format/fixed},
        scaled x ticks=false, scaled y ticks=false,
        axis lines=left,
        grid=major,
        grid style={dashed,gray!60},
        clip=false,
        legend pos=north east,
        legend style={font=\tiny,draw=none}
      ]
        
        \addplot[thick,colorcurrent,domain=0:120000,samples=200] {
          (x<=25000)*(0.1*x)
        + (x>25000 && x<=50000)*(2500 + 0.2*(x-25000))
        + (x>50000 && x<=75000)*(7500 + 0.3*(x-50000))
        + (x>75000 && x<=100000)*(15000 + 0.4*(x-75000))
        + (x>100000)*(25000 + 0.5*(x-100000))
        };
        \addlegendentry{Original}
        \addplot[thick,colorreform,domain=0:120000,samples=200] {
          (x<=20000)*0
        + (x>20000 && x<=25000)*(0.1*(x-20000))
        + (x>25000 && x<=50000)*(500 + 0.2*(x-25000))
        + (x>50000 && x<=75000)*(5500 + 0.3*(x-50000))
        + (x>75000 && x<=100000)*(13000 + 0.4*(x-75000))
        + (x>100000)*(23000 + 0.5*(x-100000))
        };
        \addlegendentry{Tax credit}
      \end{axis}
    \end{tikzpicture}
  \end{minipage}
  \caption{Example illustration of two deductibles. On the left is an input-reducing deductible that decreases the amount of taxable income by some amount $D$, in effect shifting each bracket by that amount. On the right, a tax credit equal to $D$ that nullifies the taxes paid (\euro) on the first $D$ of income.}
    \label{fig: deductibles}
\end{figure}

The second type of deductible does not shift the original brackets, but simply nullifies some part of an existing tax rule. This is illustrated in Figure \ref{fig: deductibles} on the right, again in the context of a tax bracket rule. We call these `tax-crediting deductibles'. These types of rules simply reflect a set of negative tax brackets. Dependent on what part of the tax pressure is nullified, a tax-crediting deductible is the same as an input-reducing one.\footnote{To illustrate, assume income taxes are levied at 10\% until incomes of \euro50,000 and at 20\% thereafter. An input-reducing deductible of \euro20,000 would mean that at an income before tax of \euro70,000, the absolute tax pressure is \euro5,000: income before tax of \euro70,000 minus the deductible of \euro20,000 leads to \euro50,000 which is then levied in the first bracket. A tax credit of \euro20,000 in the first bracket would lead to \euro30,000 levied at 10\% and \euro20,000 levied at 20\%, for a total pressure of \euro7,000. A tax credit in the second bracket would lead to the same \euro5,000.}

\paragraph{Miscellaneous tax rules}
Although most commonly encountered tax rules are variations of these arche-types, some are not and may violate the assumptions listed above. For example, a tax rule could be polynomial in its input. This violates the piecewise linearity assumption. There may also be tax rules that depend on multiple inputs, for example an income-dependent child benefit. For this specific example, one could view the number of dependent children as a taxpayer characteristic in which case the tax rule would constitute a conventional benefit with a single-input given a set number of children. We consider this set of miscellaneous tax rules beyond the scope of this paper although we illustrate various examples where our method can be extended to accommodate more complex setups in Section~\ref{sec:extensions}. A related approximation concerns non-refundable credits: some statutory credits cannot push an individual's tax below zero. In effect, this is a $\min(\text{credit},\,\text{gross tax})$ coupling across rules that violates Assumption~\ref{ass:additivity}. In our case study, we only treat refundable credits, but an indicator-activated cap would express the non-refundability exactly, at MILP cost.

\paragraph{The total tax code} The key idea underlying our approach is that the above assumptions allow us to succinctly describe an entire tax code as a single formula. To do so, we first introduce the concept of a `tax group'. A tax group is a set of taxpayers that share the same absolute and marginal tax pressure as long as they have the same values of the inputs. Effectively, this means that all taxpayers in a single tax group face exactly the same tax rules with the same parameters. 

By Assumption~\ref{ass:additivity} (additivity), the total tax pressure faced by a taxpayer is the sum over all individual tax rules applicable to that taxpayer. By Assumption~\ref{ass:scalar} (scalar input) each rule acts on a single input, so we may group the rules by the input they tax: let $\mathcal{R}_j$ denote the \emph{rule--input incidence set}, i.e.\ the set of rules that act on input $j$. Writing $x_{ij}$ for taxpayer $i$'s value of input $j$, the total tax code for a taxpayer $i$ in tax group $k$ can be written as

\begin{equation}
f_k(\mathbf{x}_i) = \sum_{j} \sum_{r \in \mathcal{R}_j} f_r\bigl(x_{ij};\, \Phi_{r,k}, \alpha_{r,k}\bigr) + Z_k,
\label{eq: total_tax_function}
\end{equation}
where $j$ indexes inputs, $\Phi_{r,k}$ and $\alpha_{r,k}$ are the cutoffs and rates of rule $r$ in group $k$, and $Z_k = \sum_r Z_{r,k}$ collects the lump-sum component of every rule into a single group-$k$ constant. Note that, in contrast to the per-rule indexing, the lump sum is placed \emph{outside} $f_r$ and carries only a group index. Because each $f_r$ is continuous and piecewise linear in its input by Assumption~\ref{ass:pwl}, so is their sum; the resulting representation is stated next and illustrated in Figure~\ref{fig: tax_function} for a single input.

\begin{figure}[H]
  \centering
  \definecolor{colorcurrent}{RGB}{0,114,178}   
  \definecolor{colorreform}{RGB}{213,94,0}    
  \definecolor{colorzzp}{RGB}{0,158,115}      
  \makebox[\textwidth][c]{%
    \begin{minipage}{0.43\textwidth}
      \begin{tikzpicture}
        \begin{axis}[width=0.9\linewidth,height=0.85\linewidth,font=\small,
          title={\textbf{Bracket rule}},
          ylabel={Taxes paid (\euro)},
          xmin=0,xmax=120000,ymin=0,ymax=50000,
          xtick={0,25000,50000,75000,100000},
          xticklabels={0,\euro25K,\euro50K,\euro75K,\euro100K},
          ytick={0,25000,50000},yticklabels={0,\euro25K,\euro50K},
          ticklabel style={font=\tiny,/pgf/number format/fixed},
          scaled ticks=false,
          grid=major,grid style={dashed,gray!60},axis lines=left]
          \pgfmathdeclarefunction{Tax}{1}{%
            \pgfmathparse{(#1<=25000)*(0.1*#1)+%
              (#1>25000 && #1<=50000)*(2500+0.2*(#1-25000))+%
              (#1>50000 && #1<=75000)*(7500+0.3*(#1-50000))+%
              (#1>75000 && #1<=100000)*(15000+0.4*(#1-75000))+%
              (#1>100000)*(25000+0.5*(#1-100000))}}
          \addplot[thick,black,domain=0:120000,samples=250]{Tax(x)};
          \draw[black!60,dashed] (axis cs:25000,0)    -- (axis cs:25000,50000);
          \draw[black!60,dashed] (axis cs:50000,0)    -- (axis cs:50000,50000);
          \draw[black!60,dashed] (axis cs:75000,0)    -- (axis cs:75000,50000);
          \draw[black!60,dashed] (axis cs:100000,0)   -- (axis cs:100000,50000);
        \end{axis}
      \end{tikzpicture}\vspace{0.3em}

      \begin{tikzpicture}
        \begin{axis}[width=0.9\linewidth,height=0.85\linewidth,font=\small,
          title={\textbf{Benefit rule}},
          ylabel={Taxes paid (\euro)},
          xmin=0,xmax=120000,ymin=-15000,ymax=15000,
          xtick={0,25000,50000,75000,100000},
          xticklabels={0,\euro25K,\euro50K,\euro75K,\euro100K},
          ytick={-10000,0,10000},yticklabels={-\euro10K,0,\euro10K},
          ticklabel style={font=\tiny,/pgf/number format/fixed},
          scaled ticks=false,
          grid=major,grid style={dashed,gray!60},axis lines=left,
          legend pos=north east,legend style={font=\tiny,draw=none}]
          \addplot[thick,colorcurrent]  coordinates{(0,-12000)(50000,-12000)(100000,0)(120000,0)};
          \addlegendentry{Group 1}
          \addplot[thick,colorzzp]  coordinates{(0,-8000)(75000,-8000)(100000,0)(120000,0)};
          \addlegendentry{Group 3}
          \draw[colorcurrent!60,dashed] (axis cs:50000,-15000) -- (axis cs:50000,15000);
          \draw[colorzzp!60,dashed] (axis cs:75000,-15000) -- (axis cs:75000,15000);
          \draw[black!60,dashed] (axis cs:100000,-15000) -- (axis cs:100000,15000);
        \end{axis}
      \end{tikzpicture}\vspace{0.3em}

      \begin{tikzpicture}
        \begin{axis}[width=0.9\linewidth,height=0.85\linewidth,font=\small,
          title={\textbf{Tax-crediting deductible}},
          xlabel={Income before tax (\euro)},ylabel={Taxes paid (\euro)},
          xmin=0,xmax=120000,ymin=-15000,ymax=15000,
          xtick={0,25000,50000,75000,100000},
          xticklabels={0,\euro25K,\euro50K,\euro75K,\euro100K},
          ytick={-10000,0,10000},yticklabels={-\euro10K,0,\euro10K},
          ticklabel style={font=\tiny,/pgf/number format/fixed},
          scaled ticks=false,
          grid=major,grid style={dashed,gray!60},axis lines=left,
          legend pos=south east,legend style={font=\tiny,draw=none}]
          \addplot[thick,colorcurrent] coordinates{(0,0)(120000,0)};
          \addlegendentry{Group 1}
          \addplot[thick,colorreform] coordinates{(0,0)(25000,-2500)(120000,-2500)};
          \addlegendentry{Group 2}
          \draw[colorreform!60,dashed] (axis cs:25000,-15000) -- (axis cs:25000,15000);
        \end{axis}
      \end{tikzpicture}
    \end{minipage}%
    \hspace{0.5cm}%
    \begin{minipage}{0.04\textwidth}
      \centering
      \begin{tikzpicture}[baseline]
        \draw[decorate,decoration={brace,amplitude=6pt},very thick]
          (0,7) -- (0,-7);
      \end{tikzpicture}
    \end{minipage}%
    \begin{minipage}{0.53\textwidth}
      \centering
      \begin{tikzpicture}
        \begin{axis}[width=\linewidth,height=0.8\linewidth,font=\small,
          title={\textbf{Combined tax function}},
          xlabel={Income before tax (\euro)},ylabel={Taxes paid (\euro)},
          xmin=0,xmax=120000,ymin=-30000,ymax=50000,
          xtick={0,25000,50000,75000,100000},
          xticklabels={0,\euro25K,\euro50K,\euro75K,\euro100K},
          ytick={-20000,0,25000,50000},yticklabels={-\euro20K,0,\euro25K,\euro50K},
          ticklabel style={font=\tiny,/pgf/number format/fixed},
          scaled ticks=false,
          grid=major,grid style={dashed,gray!60},axis lines=left,
          legend pos=north west,legend style={font=\tiny,draw=none}]
          \pgfmathdeclarefunction{Tax}{1}{%
            \pgfmathparse{(#1<=25000)*(0.1*#1)+%
              (#1>25000 && #1<=50000)*(2500+0.2*(#1-25000))+%
              (#1>50000 && #1<=75000)*(7500+0.3*(#1-50000))+%
              (#1>75000 && #1<=100000)*(15000+0.4*(#1-75000))+%
              (#1>100000)*(25000+0.5*(#1-100000))}}
          \pgfmathdeclarefunction{FlatOne}{1}{\pgfmathparse{0}}
          \pgfmathdeclarefunction{FlatTwo}{1}{%
            \pgfmathparse{(#1<=25000)*(-0.1*#1)+(#1>25000)*(-2500)}}
          \pgfmathdeclarefunction{IncOne}{1}{%
            \pgfmathparse{%
              (#1<=50000)*(-12000)+%
              (#1>50000 && #1<=100000)*(-12000*(1-(#1-50000)/50000))}}
          \pgfmathdeclarefunction{IncThr}{1}{%
            \pgfmathparse{%
              (#1<=75000)*(-8000)+%
              (#1>75000 && #1<=100000)*(-8000*(1-(#1-75000)/25000))}}
          \addplot[thick,colorcurrent,  domain=0:120000,samples=250]{Tax(x)+FlatOne(x)+IncOne(x)};
          \addlegendentry{Group 1}
          \addplot[thick,colorreform,domain=0:120000,samples=250]{Tax(x)+FlatTwo(x)};
          \addlegendentry{Group 2}
          \addplot[thick,colorzzp,  domain=0:120000,samples=250]{Tax(x)+IncThr(x)};
          \addlegendentry{Group 3}
          \draw[black!60,dashed] (axis cs:25000,-30000) -- (axis cs:25000,50000);
          \draw[black!60,dashed] (axis cs:50000,-30000) -- (axis cs:50000,50000);
          \draw[black!60,dashed] (axis cs:75000,-30000) -- (axis cs:75000,50000);
          \draw[black!60,dashed] (axis cs:100000,-30000)-- (axis cs:100000,50000);
        \end{axis}
      \end{tikzpicture}
    \end{minipage}
  }
  \caption{Illustration of how three tax rules lead to a single piecewise linear function. The top inset on the left illustrates a bracket rule. The middle inset on the left illustrates a group-specific and income-dependent benefit. The bottom inset on the left illustrates a tax crediting deductible. Per rule, shared cutoff points are denoted with black dashed lines. Group-specific ones are denoted with that group's color.}
  \label{fig: tax_function}
\end{figure}

\begin{proposition}[Representation]\label{prop:representation}
Under Assumptions~\ref{ass:additivity}--\ref{ass:hetero}, the total tax liability of any taxpayer $i$ in tax group $k$ is the separable, continuous, piecewise-linear function
\begin{equation}
f_k(\mathbf{x}_i) = \sum_{j} \sum_{b \in \mathcal{B}_{j,k}} \alpha_{j,b,k}\, g_b\bigl(x_{ij};\, \Phi_{j,k}\bigr) + Z_k,
\label{eq: representation}
\end{equation}
where, for each input $j$, the support $\Phi_{j,k} = \bigcup_{r \in \mathcal{R}_j} \Phi_{r,k}$ is the union of the cutoffs of all group-$k$ rules acting on input $j$, $\mathcal{B}_{j,k}$ indexes the brackets it induces, $\alpha_{j,b,k}$ is the consolidated marginal rate on bracket $b$, $g_b(\cdot;\Phi_{j,k})$ is the bracketing function of Definition~\ref{def:bracketing}, and $Z_k$ is the group-$k$ lump sum. In particular, $f_k$ is piecewise linear in each input, with breakpoints contained in the union of all rule cutoffs on that input.
\end{proposition}

The proof is given in~\ref{sec: proofs}. Equation~\eqref{eq: representation} is central to our approach: it shows that the entire statutory code, however many rules it contains, reduces per tax group to one rate vector per input over a known breakpoint set, plus a lump sum.

\section{Tax reform as constrained optimization}
\label{sec:reform}

Viewing the tax code as one piecewise-linear function per tax group exposes two sets of quantities that determine the actual statutory code: the cutoff points that divide each input into brackets, which form the system's support $\Phi$, and the rates levied on each bracket together with any lump-sum transfers, which we collect as the system's rates $\mathcal{A}$. A reform is a new set of rates $\mathcal{A}^\star$ and, if desired, a new support $\Phi'$ (Section~\ref{sec:cases}). Throughout this paper we use \texttt{Gurobi} as our numerical solver \citep{gurobi}. Our open source \texttt{TaxSolver} package supports other solvers, as well.

\subsection{Recovering the current system}
\label{sec:recovery}

To illustrate tax design as a constrained optimization problem, we start by showing that we can recover an existing system's rates from data. Assume we have a dataset on taxpayers containing tax-relevant inputs (e.g.\ gross income), taxpayer characteristics (e.g.\ fiscal status), and observed tax pressures under the current system, and that we know the support of the system. We can then recover the current rates by solving the feasibility problem

\begin{equation}
\begin{aligned}
    \min_{\mathcal{A}} \quad & 0 \\
    \text{s.t.} \quad & f(x_i, \mathcal{A})|_\Phi = y'_i, \quad \forall i \in \{1, \dots, n\},
\end{aligned}
\label{eq: recovery}
\end{equation}
where the rates $\mathcal{A}$ are the decision variables and the constraints require every taxpayer's absolute tax pressure, $f(x_i, \mathcal{A})$, in the `reformed' system to equal their pressure under the current system, $y_i'$. With a constant objective and linear equality constraints, this is a linear feasibility problem (a linear program, or LP). Whether its solution pins down the current system uniquely is an identifiability question:

\begin{proposition}[Recovery and identifiability]\label{prop:recovery}
Stack the recovery equality constraints as $\mathbf{y}' = X\boldsymbol{\theta}$, where $\boldsymbol{\theta} \in \mathbb{R}^{P}$ collects the free parameters of the system (the bracket rates $\alpha_{j,b,k}$ and the lump sums $Z_k$) and each row of the design matrix $X \in \mathbb{R}^{N \times P}$ holds the corresponding bracketed inputs $g_b(x_{ij};\Phi_{j,k})$ and group indicators for one taxpayer. If $X$ has full column rank, the recovery problem has a unique solution, equal to the true parameters. If $X$ is column-rank-deficient, the feasible set is the affine subspace $\{\boldsymbol{\theta}^\star + v : v \in \ker X\}$, and every parameterization in it reproduces the observed liabilities exactly.
\end{proposition}

Full column rank requires sufficient interior income variation in every bracket-by-group cell. This condition can fail in two instructive ways. First, a bracket that every taxpayer in a group traverses completely contributes a \emph{constant} column, equal to its width $\phi_b - \phi_{b-1}$ that is collinear with that group's lump-sum column. The data then cannot separate the bracket's rate from the lump sum. Second, a bracket that no taxpayer reaches contributes a \emph{zero} column, leaving its rate unidentified. This is the familiar non-uniqueness of inverse problems.

Loosening the equality constraints enlarges the feasible set and yields alternative reforms $\mathcal{A}^\star$.

\subsection{Setting fiscal guarantees}
\label{sec:guarantees}

To cap alternative solutions to the system's current rates when relaxing the equality constraints, we introduce a set of constraints, or \emph{fiscal guarantees}, that encode the theoretical and practical considerations a reform should satisfy.

\paragraph{Income constraints} The most important fiscal guarantee bounds how far a taxpayer's tax burden may move under the new system. We state it on \emph{net income}, $n_i$, rather than on tax paid. Let
\begin{equation*}
n_i(\mathcal{A}) = x_{i,\text{income}} - f(x_i, \mathcal{A}), \qquad n_i' = x_{i,\text{income}} - y_i'
\end{equation*}
denote net income under the reform and under the status quo. The income constraint is the two-sided guarantee
\begin{equation}
(1 - \underline{\epsilon}_i)\, n_i' \;\le\; n_i(\mathcal{A}) \;\le\; (1 + \overline{\epsilon}_i)\, n_i', \qquad i = 1, \dots, n,
\label{eq: income_constraint}
\end{equation}
where $\underline{\epsilon}_i, \overline{\epsilon}_i \ge 0$ are the largest permitted proportional decrease and increase in net income for taxpayer $i$ and may be set per taxpayer or per group; tight constraints set $\underline{\epsilon}_i = \overline{\epsilon}_i = 0$ and recover the status quo. Stating the guarantee on net income, which is typically non-negative, rather than on tax paid avoids a sign problem: for net benefit recipients the status-quo liability $y_i'$ is negative, so a proportional bound written directly on $f(x_i, \mathcal{A})$ would reverse direction. The constraint~\eqref{eq: income_constraint} is proportional, but absolute bounds like a poverty floor can be imposed by replacing the multiplicative factors with additive ones. The applied reforms in this paper impose only the floor ($\overline{\epsilon}_i = \infty$), so individual windfalls are disciplined only in aggregate through the budget band; where large idiosyncratic gains are themselves politically salient, the upper bound can be applied. For households with non-positive status-quo net income the multiplicative floor remains valid and, if anything, conservative: with $n'_h \le 0$, \eqref{eq: income_constraint} requires $n_h(\mathcal{A}) \ge (1-\underline{\epsilon})\,n'_h \ge n'_h$, so no loss is admitted.

\paragraph{Marginal constraints} Another type of fiscal guarantee could be to determine the marginal pressure experienced by taxpayers.  Marginal pressure curves are a typical feature of optimal taxation theory as prohibitively high marginal pressure can reduce labor participation incentives. We call these `marginal constraints'. When a single rule governs a single input, the constraint reduces to a direct bound on that rule's own rate:

\begin{equation}
\alpha_b \leq \tau_{\text{max}}, \quad \forall b,
\label{eq: marginal_constraint}
\end{equation}
where $\tau_{\text{max}}$ is the maximum allowed marginal rate, and both lower and upper bounds can be set.

Note that, several rules can act on inputs that co-move with the same euro of a given earner's own earnings, for instance the earner's personal income and a household-level input that both change with it (illustrated in Section~\ref{sec: case_4_dutch}).

\paragraph{Budget constraints} A third type of fiscal guarantee can be used to limit the shock to government revenue due to reform. We call these `budget constraints' and they can be implemented analogous with the income constraints introduced above, as the total tax revenue is simply the sum over all individuals' taxes paid:

\begin{equation}
\sum_i w_i\,[y'_i - f(x_i, \mathcal{A})] \leq C.
\label{eq: budget_constraint}
\end{equation}
Here, $C \in \mathbb{R}$ is the maximum allowed loss in tax revenue, with $w_i$ the sampling weight of taxpayer $i$; the symmetric band $\bigl|\sum_i w_i\,[y'_i - f(x_i, \mathcal{A})]\bigr| \le C$, used in the Dutch case study of Section~\ref{sec:cases}, bounds windfalls as well as losses.

\paragraph{Distributional constraints and objectives} The income constraint~\eqref{eq: income_constraint} is stated per taxpayer, but distributional targets can enter the same problem class through aggregation. A poverty-gap term is the canonical example. Introducing for each taxpayer a shortfall variable $s_i \geq 0$ with $s_i \geq z_i - n_i(\mathcal{A})$, where $z_i$ is the applicable poverty line (data, per taxpayer or household), the weighted aggregate gap $\sum_i w_i s_i$ can be bounded as a constraint or minimized as an objective. This costs one continuous variable and one linear row per taxpayer and leaves the formulation in the LP class. The same construction linearizes any measure built from lower-tail shortfalls against fixed thresholds; rank-based inequality measures such as the Gini coefficient are not linear in the rates, but ordered-weighted proxies (e.g.\ minimizing the sum of the $k$ largest shortfalls) admit standard LP reformulations. The household income floors applied throughout this paper are the special case in which each shortfall is forced to zero at threshold $z_h = 0.95\,n'_h$. Figure~\ref{fig: case_nl_incidence} of the Dutch case study illustrates this.

Many fiscal guarantees can also feature as objective functions. For example, minimizing the revenue-loss expression on the left of~\eqref{eq: budget_constraint} finds the cheapest reform given the constraints and keeps the problem in the LP class. More complex objectives, like system complexity, might introduce the need for approximate or iterative methods as we will illustrate below.

\subsection{Setting the support, tax groups and tax rules for a reform}
\label{sec:supportchoice}

During reform, the same support and tax groups as the old system can be adopted. However, changes to either can be implemented as well. For example, limiting the support of the system can be utilized to mechanically simplify the resulting solution, for example by removing cutoff points from the support. Conversely, new cutoffs can be added to allow for new tax brackets. The same holds for the amount of tax groups that are available to reform the system. Rather than using the same tax groups as present in an existing system, policymakers can decide to reduce the extent to which taxpayers might experience different tax rules by removing tax groups or create new ones.

Finally, it might be politically desirable to keep certain tax rules and reform the system net of these rules. This can also be implemented by simply setting certain decision variables to designated values instead of making them eligible for reform.

\subsection{Extensions}
\label{sec:extensions}


The formulation so far yields linear programs. We now describe five extensions that broaden its reach: a cardinality objective for simplicity, endogenous bracketing, lexicographic multi-objective reform, multi-step reform and behavioral responses. We make explicit where each moves the problem beyond the LP class into mixed-integer linear or quadratic territory. The first four are also formalized here. Behavioral responses are treated in Section~\ref{sec: case_3_behavioral}.

\paragraph{Simplification as a weighted-cardinality objective} Above, tax code simplicity was treated \emph{mechanically} by pruning the support of the tax groups before solving. Simplicity can instead be made an explicit objective using a big-$M$ approach. Associate with each candidate bracket rate $\alpha_b$ a binary activation variable $z_b \in \{0,1\}$ and link the two with
\begin{equation}
-\tau_{\max}\, z_b \;\le\; \alpha_b \;\le\; \tau_{\max}\, z_b, \qquad \forall b,
\label{eq: activation}
\end{equation}
so that $z_b = 0$ forces $\alpha_b = 0$ (the bracket is inactive) while $z_b = 1$ leaves $\alpha_b$ free within the admissible box $[\underline{\alpha}_b, \overline{\alpha}_b]$ (whose upper end is capped by the marginal guarantee~\eqref{eq: marginal_constraint}). Lump sums $Z$ enter the count through the same link, with $\tau_{\max}$ replaced by the monetary cap placed on each benefit. Minimizing the weighted count of active rules,
\begin{equation}
\min_{\mathcal{A},\,z}\ \sum_b w_b\, z_b,
\label{eq: cardinality_obj}
\end{equation}
penalizes complexity, with weights $w_b$ set to the administrative burden each rule carries. This is a weighted-cardinality ($\ell_0$-type) objective of the kind that modern mixed-integer optimization solves at scale~\citep{bertsimas2016best}, and it turns the reform into a mixed-integer linear program (MILP).

The same approach can be used to retain or scale an existing complex rule: writing $p_{i,r}$ for the fixed pressure that legacy rule $r$ exerts on taxpayer $i$ under current law, the solver may admit a scaled copy $s_r\, p_{i,r}$ subject to
\begin{equation}
0.9\, z_r \;\le\; s_r \;\le\; 1.1\, z_r, \qquad z_r \in \{0,1\},
\label{eq: reuse_scaling}
\end{equation}
so a legacy rule can be dropped ($z_r = 0 \Rightarrow s_r = 0$), kept as is ($s_r = 1$), or rescaled within $[0.9, 1.1]$, while its activation $z_r$ can still enter~\eqref{eq: cardinality_obj} with a legacy weight.

\paragraph{Dynamic bracketing} The support itself can be treated as a decision variable as well. Fix a dense candidate support $\bar\Phi = \{\phi_0 < \cdots < \phi_B\}$, let the
solver assign a rate $\alpha_b \in [\underline\alpha, \overline\alpha]$ to each candidate interval, and introduce rate-change indicators $d_b \in \{0,1\}$, $b = 1,\dots,B-1$, with
\begin{equation}
-M d_b \le \alpha_{b+1} - \alpha_b \le M d_b, \quad
\sum_{b=1}^{B-1} d_b \le K-1, \qquad M = \overline\alpha - \underline\alpha .
\label{eq: dyn_bracket}
\end{equation}

The rate may change at $\phi_b$ only if $d_b = 1$, so at most $K-1$ changes collapse the candidate grid into a schedule with at most $K$ brackets, whose cutoffs the solver selects along with the rates.\footnote{$M = \tau_{\max}$ would be valid only if all rates shared an interval of that width; signed consolidated rates admit differences up to the full width of the rate box.} Because bracket membership is evaluated on the fixed grid, liabilities stay linear in $\alpha$; continuous cutoffs would multiply rates by cutoffs and leave the MILP class. This is an $\ell_0$ restriction on adjacent rate differences~\citep{natarajan1995sparse, toriello2012fitting}.

\paragraph{Multi-objective reform via lexicographic optimization} A reform often pursues several goals, for instance minimal revenue loss \emph{and} minimal complexity. We treat them lexicographically. The first stage minimizes revenue loss subject to the fiscal guarantees and records the optimal loss $L^\star$. The second stage minimizes the cardinality objective~\eqref{eq: cardinality_obj} over the same feasible set, subject to the tolerance constraint that revenue loss not exceed $L^\star$ by more than a slack $\beta$, so that a bounded amount of revenue can be spent to buy simplicity. The Dutch study of Section~\ref{sec:cases} runs exactly this procedure with $\beta$ equal to a further $1\%$ of current revenue.

\paragraph{Multi-step reform} Finally, step-wise routines decompose a large reform into stages: a first reform that meets an initial set of criteria becomes the status quo for a second that adds further ones. This is useful when a reform must clear sequential political hurdles, each with its own guarantees.

The above extensions all trade tractability for flexibility. The cardinality and dynamic-bracketing objectives add binary variables and turn the LP into an MILP. Including behavioral responses, which we discuss below, adds a rate-by-base product and yields a nonconvex MIQCP. Both classes remain tractable as the behavioral example and the Dutch case study of Section~\ref{sec:cases} demonstrate.

\subsection{Incorporating behavioral responses}
\label{sec: case_3_behavioral}

Reforms that change marginal rates also change taxpayer behavior, and a framework that ignores this can misstate a reform's revenue cost. Let $\delta \in \mathbb{R}$ denote the elasticity of taxable income: the proportional change in reported income per proportional change in the net-of-tax rate $1 - \tau$~\citep{GruberSaez2002_ETI}. We take $\delta$ common across taxpayers for simplicity, though taxpayer-specific $\delta_i$ enter identically. Linearizing the log-linear specification $\Delta\log x_i = \delta\,\Delta\log(1-\tau_i)$ gives
\begin{equation}
x_i^{\text{new}} = x_i\left(1 + \delta\,\frac{\tau_i^{\text{sq}} - \tau_i^{\text{new}}}{1 - \tau_i^{\text{sq}}}\right),
\label{eq: behavioral_income}
\end{equation}
where $x_i{^\text{new}}$ is taxpayer $i$'s reported new income under the reform, $\tau_i^{\text{sq}}$ is taxpayer $i$'s active marginal rate under current law (data) and $\tau_i^{\text{new}}$ is the active marginal rate under the reform. A higher reformed rate lowers reported income, and vice-versa. Income guarantees apply to net income at $x_i^{\text{new}}$ but remain indexed to status-quo net income $n_i'$. Because $\tau_i^{\text{sq}}$ is data,~\eqref{eq: behavioral_income} is linear in $\tau_i^{\text{new}}$.

\paragraph{Bracket stability} The rate $\tau_i^{\text{new}}$ is the rate of whichever bracket $x_i^{\text{new}}$ falls into, so the response in~\eqref{eq: behavioral_income} can in principle push a taxpayer across a cutoff and make $\tau_i^{\text{new}}$ itself a function of $x_i^{\text{new}}$. We assume each taxpayer remains in their status-quo bracket, so $\tau_i^{\text{new}}$ is the (decision-dependent) rate of a \emph{fixed} bracket. The assumption is likely innocuous for smooth schedules, but it can be relaxed by adding bracket-assignment binaries at the cost of a larger model. Empirically, taxpayers whose response would carry them across a cutoff instead bunch at it~\citep{Saez2010_Bunching, Kleven2016_Bunching}. In the rate-raising reforms below, would-be crossers move \emph{down}, so bracket stability overstates income loss and revenue erosion; the bias reverses for rate cuts.

Even under bracket stability, substituting~\eqref{eq: behavioral_income} into the income and budget constraints introduces the product of two decision variables: the rate $\tau_i^{\text{new}}$ and the income response it induces. Its continuous relaxation becomes a nonconvex quadratically constrained program (QCQP); together with the binary activation variables of Section~\ref{sec:extensions} the program is a nonconvex MIQCP. Modern solvers handle this class directly at the scale considered here, and the direct solve is our primary method. For backends without quadratic support, Algorithm~\ref{alg:behavioral} provides an LP-only alternative: hold incomes fixed, solve the resulting linear reform program for the rates, update incomes through~\eqref{eq: behavioral_income} with a damped step, and repeat until incomes stop moving. There is no certificate of optimality with this approach, but we implement both in our illustrative example with comparable results, below.

\begin{algorithm}[!t]
\caption{Fixed-point reform with behavioral responses}
\label{alg:behavioral}
\begin{algorithmic}[1]
\State \textbf{input:} data $\{x_i, y_i', \tau_i^{\text{sq}}\}$, elasticity $\delta$, damping $\lambda \in (0, 1]$, tolerance $\eta$, fiscal guarantees and objective
\State initialize $x_i^{(0)} \gets x_i$ for all $i$; \quad $t \gets 0$
\Repeat
    \State solve the \emph{linear} reform program with incomes fixed at $x_i^{(t)}$, obtaining rates $\mathcal{A}^{(t)}$ and active marginal rates $\tau_i^{\text{new},(t)}$
    \State $\tilde{x}_i \gets x_i\bigl(1 + \delta\,(\tau_i^{\text{sq}} - \tau_i^{\text{new},(t)})/(1 - \tau_i^{\text{sq}})\bigr)$ for all $i$ \Comment{undamped target}
    \State $x_i^{(t+1)} \gets (1 - \lambda)\, x_i^{(t)} + \lambda\, \tilde{x}_i$ for all $i$
    \State $t \gets t + 1$
\Until{$\max_i \lvert \tilde{x}_i - x_i^{(t-1)} \rvert \le \eta$}
\State \textbf{return} rates $\mathcal{A}^{(t-1)}$
\end{algorithmic}
\end{algorithm}

\subsection{Infeasibility certification}
\label{sec:iis}

A distinctive feature of casting reform as an optimization problem is that the model can \emph{prove} that no reform meeting a given set of guarantees exists. When the guarantees conflict, the solver reports infeasibility, and modern solvers can additionally return an \emph{irreducible infeasible subsystem} (IIS): a subset of constraints that is itself infeasible but becomes feasible if any single member is removed~\citep{chinneck2008feasibility}. For linear and mixed-integer programs an IIS is isolated by deletion and additive filtering~\citep{chinneck1991locating}, and is exposed directly by the solver we use.

The IIS makes infeasibility results actionable for policymakers. Every constraint in our formulation corresponds to a named guarantee, so an IIS translates into a plain statement of which commitments cannot hold at once. We illustrate this ability in our Dutch case study, below.

\subsection{Illustrative examples}
\label{sec: case_1_single}
\label{sec: case_2_multiple}

In what follows, we build intuition on a few small instances before turning to the national-scale code in Section~\ref{sec: case_4_dutch}. All toy solves use $N=1{,}000$ taxpayers, but to avoid cluttering our scatter plots show the first $100$. All computational results, including model dimensions, runtimes, hardware, and optimality gaps are reported in the Supplementary Material.

\paragraph{A single progressive rule}
Consider the single progressive bracket rule of Figure~\ref{fig: brackets} (left), which brackets income at \euro25,000, \euro50,000, \euro75,000 and \euro100,000 at rates of 10\%, 20\%, 30\%, 40\% and 50\%. Incomes are spread across the bracket support (Figure~\ref{fig: case_1_combined}a); Jude (\euro52,000, tax \euro8,100) and Laila (\euro120,000, tax \euro35,000) are highlighted as two representative rows. Bracketing income along the support writes each liability as a linear function of the rate vector. For example,
\begin{equation*}
    y_\text{Jude} = 25\,000\,\alpha_1 + 25\,000\,\alpha_2 + 2\,000\,\alpha_3 + 0\,\alpha_4 + 0\,\alpha_5,
\end{equation*}
\begin{equation*}
    y_\text{Laila} = 25\,000\,\alpha_1 + 25\,000\,\alpha_2 + 25\,000\,\alpha_3 + 25\,000\,\alpha_4 + 20\,000\,\alpha_5,
\end{equation*}
with the rates $\alpha$ as decision variables. Tight income constraints, requiring $y_\text{Jude}=8\,100$ and $y_\text{Laila}=35\,000$, admit only the status quo and recover the current rates exactly (Figure~\ref{fig: case_1_combined}a). Widening the constraints opens up alternatives: targeting taxpayers below \euro70,000 with a net-income increase of at least 5\%, allowing all others decreases of up to 10\%, and minimizing revenue loss yields the reform in Figure~\ref{fig: case_1_combined}b. With only lower bounds active, minimizing revenue loss is revenue-\emph{maximizing} subject to the guarantees, which is why the solver pushes the top rates to 64\% and collects 7.4\% more revenue than the status quo.

\begin{figure}[!t]
    \centering
    \includegraphics[width=\linewidth]{figs/case_1a_combined.png}
    \caption{(a) Income before tax and taxes paid of 100 of 1{,}000 taxpayers, including Jude and Laila (highlighted in orange). The dashed line illustrates rates when setting tight income constraints. (b) Illustration of loose income constraints, where taxpayers with incomes below \euro70,000 are forced to have an increase in net income of at least 5\% and all other taxpayers can face decreases of up to 10\%. The dashed line shows reform rates.}
    \label{fig: case_1_combined}
\end{figure}

\paragraph{Adding rules}
We now enrich the system with two further rules: a \euro1,500 healthcare benefit nullified between incomes of \euro30,000 and \euro40,000, and an \euro800-per-child benefit.  Under tight constraints on the same instance, the solver reproduces every liability and returns the generating schedule (Figure~\ref{fig: case_2}a). The same reform can then be combined with additional commitments: a flat 60\% marginal-pressure cap, the child benefit held fixed (outside the reform), and the healthcare nullification removed from the support. These enter as a marginal constraint, as fixed decision variables, and as a support restriction, respectively, and produce the reform in Figure~\ref{fig: case_2}b.

\begin{figure}[!t]
    \centering
    \includegraphics[width=\linewidth]{figs/case_1b_combined.png}
    \caption{The dashed lines indicate reforms and the solid blue lines the current system. Panel (a) plots the solver's returned recovery schedule under tight equality constraints. Panel (b) enacts the same reform as Figure~\ref{fig: case_1_combined}b under a flat 60\% marginal-pressure cap, with the child benefit held fixed outside the reform and the support of income before tax limited to five brackets.}
    \label{fig: case_2}
\end{figure}

\begin{figure}[!t]
    \centering
    \includegraphics[width=\textwidth]{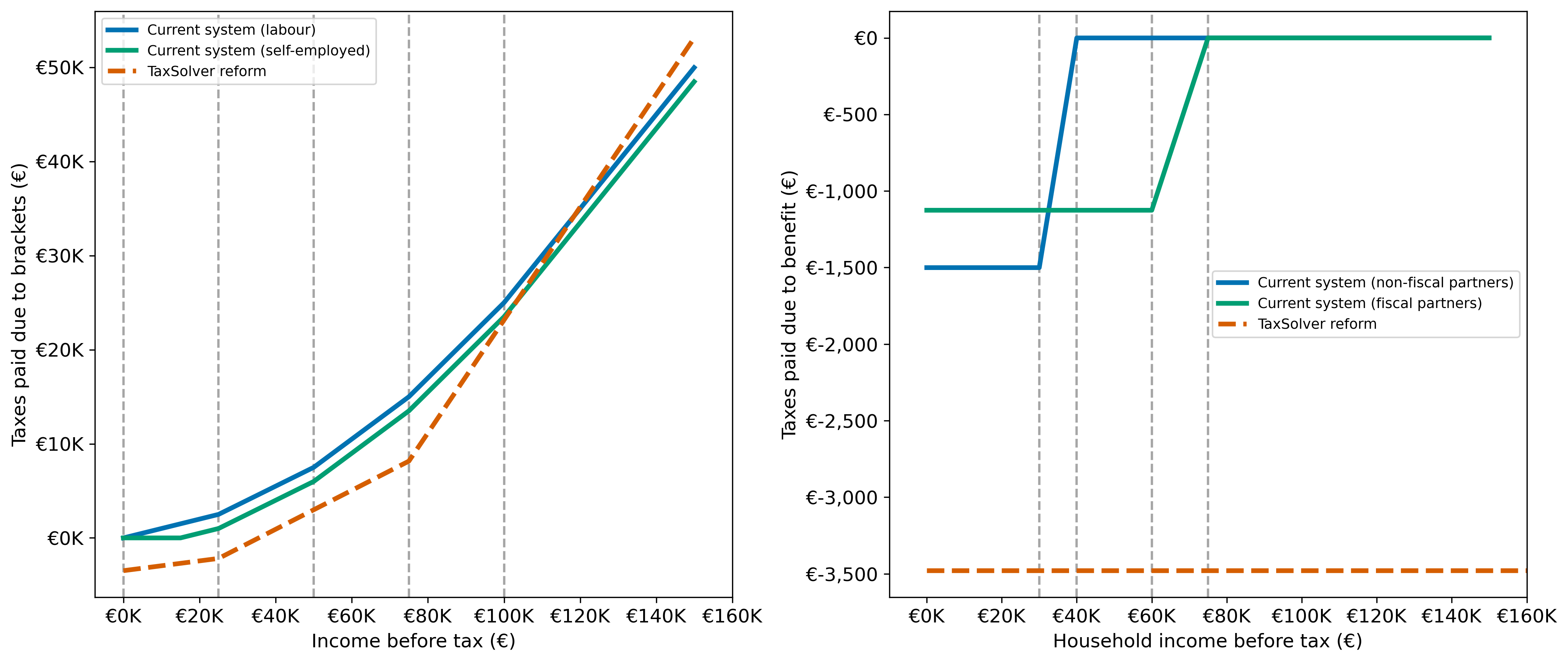}
    \caption{The above illustrates reforms realizing income support of at least 5\% for households earning income before tax below \euro85,000 and removing separate brackets for the self-employed versus employees. The childcare benefit in the reform is set at \euro713.}
    \label{fig: case_3_reform}
\end{figure}

\paragraph{Adding tax groups}
Finally we add two taxpayer characteristics that induce tax groups. Fiscal-partnership status splits the healthcare benefit: \euro2,250 for partnered households (nullified as household income rises from \euro60,000 to \euro75,000) and \euro1,500 for singles (nullified from \euro30,000 to \euro40,000). Employment status adds a self-employment credit exempting the first \euro15,000 of personal income. Recovery extends as before, now including household income as an input; the universal income schedule is shared across groups, with group-specific add-on rules for healthcare, self-employment, and lump sums. We omit its figure and proceed to the reform. Setting income constraints at the household level, forcing a 5\% net-income increase for households below a combined \euro85,000, and simplifying by removing the healthcare brackets and all group-specific dependence yields the reform in Figure~\ref{fig: case_3_reform}.

\begin{figure}[!t]
    \centering
\includegraphics[width=\linewidth]{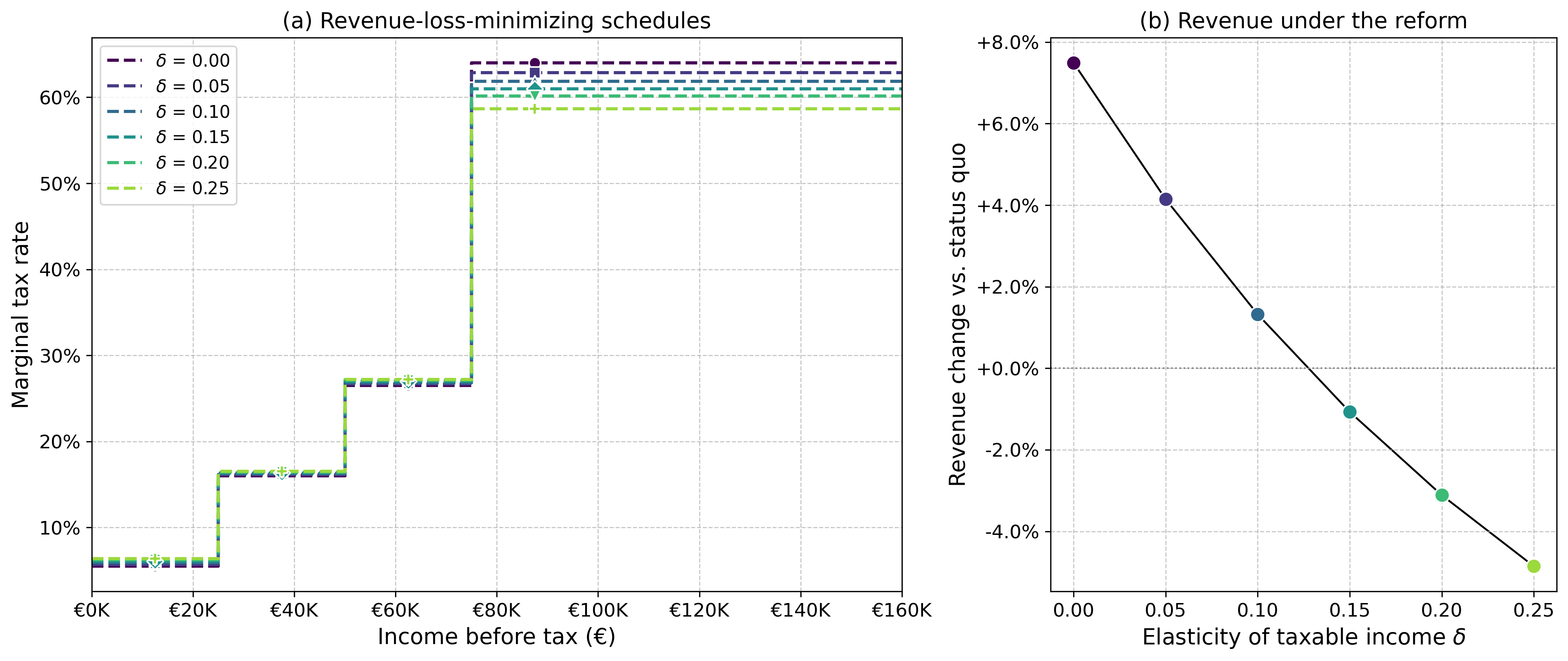}
    \caption{Behavioral responses on the single-rule instance. Left: revenue-loss-minimizing rate schedules at elasticities $\delta \in [0, 0.25]$. The income guarantees pin the lower brackets, so the response concentrates in the top rate. Right: revenue under the reform relative to the status quo. Behavioral erosion lowers the revenue the cheapest conforming schedule can collect, from $7.5\%$ above status-quo revenue at $\delta = 0$ to $4.9\%$ below it at $\delta = 0.25$.}
    \label{fig: case_3_behavioral}
\end{figure}

\paragraph{Behavioral effects}
Before moving to our real-world case study, we illustrate both the direct solve as well as the fixed-point approach from Algorithm 1 on an example that includes behavioral effects. We use the single-group, single-rule instance above ($N=1{,}000$) and sweep elasticities $\delta \in [0, 0.25]$ in steps of 0.05. Each solve minimizes revenue loss subject to the same income guarantees as the earlier reform examples. Figure~\ref{fig: case_3_behavioral} shows the effect of the behavioral response: revenue under the cheapest conforming reform falls monotonically with the elasticity, from $7.5\%$ above status-quo revenue at $\delta = 0$ to $4.9\%$ below it at $\delta = 0.25$. The damped iteration ($\lambda = 0.5$) converges within at most 14 outer iterations for every $\delta$ and agrees with the direct nonconvex solve to within $0.03$ percentage points on every rate up to $\delta = 0.20$, and to within $0.781$ percentage points at $\delta = 0.25$ (see the convergence table in the Supplementary Material).\footnote{Algorithm~\ref{alg:behavioral} has no global convergence guarantee: optimal rates change only when an income crosses a bracket cutoff, so the undamped map can cycle at larger $\delta$. Damping restores convergence in every run reported here, but the income fixed point need not coincide with the coupled direct solve. For the $\delta=0.25$ the rate gap is confined to the top bracket and corresponds to only $0.22\%$ of status-quo revenue.}

\section{The Dutch income tax code}
\label{sec:cases}
\label{sec: case_4_dutch}

This section applies the framework to a real-world tax system. We reproduced the main body of tax rules in the Dutch income tax code (shown in Table \ref{table: rule_topics}). We then generated a simulated dataset covering a large number ($N=13,500$) of Dutch taxpayers and households.\footnote{One consequential rule, the mortgage-interest deduction, is excluded for lack of public data on mortgage deduction (see the Supplementary Material).} This case study is representative of applying our method in a real-world setting with access to microdata. However, it does not exactly constitute a set of realistic reform proposals for the Netherlands, as the input data is a synthetic set of households based on publicly available aggregated statistics (see the Supplementary Material).

\begin{figure}[!t]
    \centering
    \includegraphics[width=\textwidth]{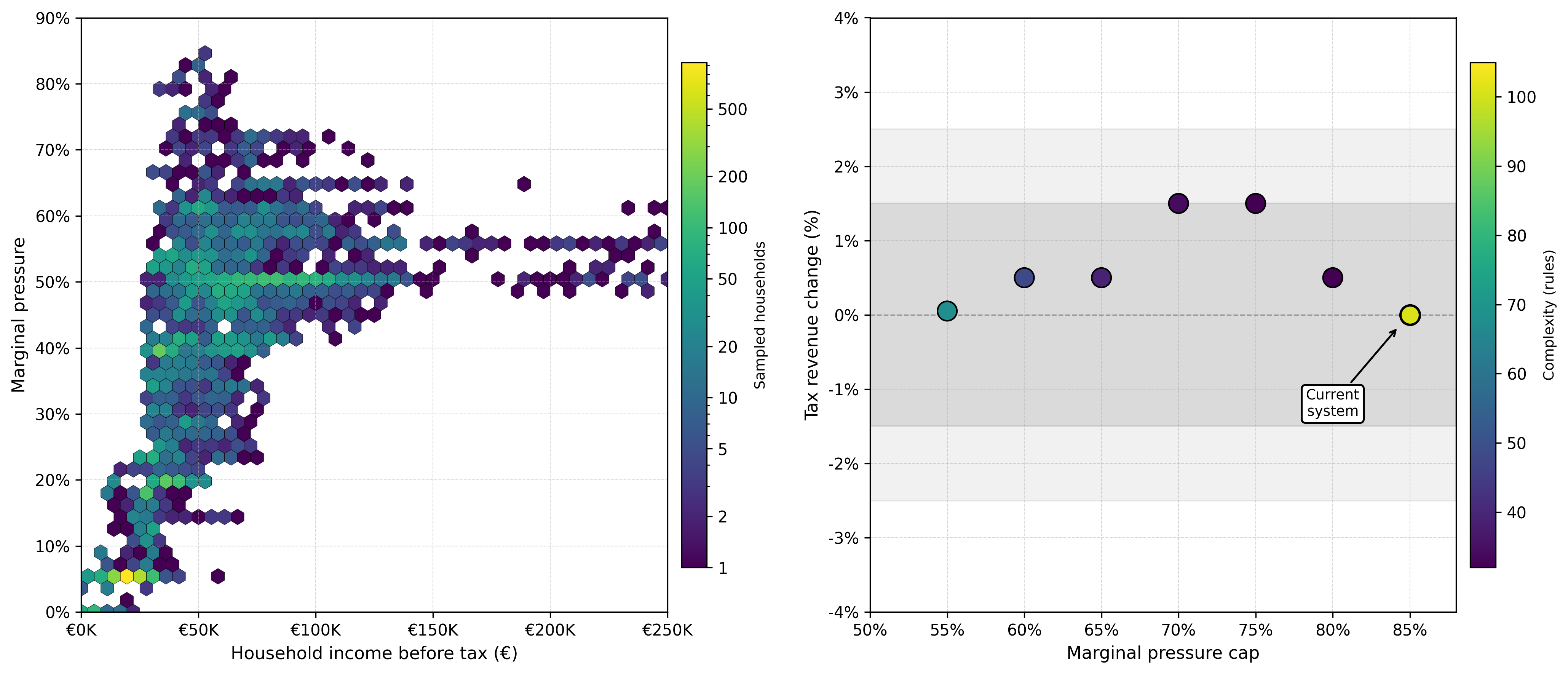}
    \caption{Left: marginal pressure faced by households in the Dutch system at different values of income before tax. Right: loss in tax revenue as a function of the cap on marginal pressure per reform. The color coding illustrates the weighted rule count of each reform under the complexity scheme of Section~\ref{sec:cases}. The dark grey band illustrates the original +1.5\% and -1.5\% budget constraint and the light grey band the further +1\% and -1\% for the second objective.}
    \label{fig: case_4_intro_comp}
\end{figure}

The Netherlands is a case in point where many ad-hoc reforms have led to a complex code with unintended characteristics, as is illustrated by the marginal pressure experienced by taxpayers (Figure \ref{fig: case_4_intro_comp}, left). Many low- and middle-income taxpayers face marginal pressure upwards of 80\% due to multiple income-dependent benefits simultaneously tapering off around similar income points. Such spikes in marginal pressure are contrary to what optimal taxation theory prescribes~\citep{Saez2001_Elasticities, diamond1998optimal} and are a recurring topic of debate in Dutch Parliament~\citep{CPB2020KansrijkBelastingbeleid,MinFin2024Miljoenennota}.

\begin{sidewaystable}[p]   
  \centering
  \resizebox{\textheight}{!}{%
    \begin{tabular}{lcccccc}
    \toprule
    {Rule topic} &
    \multicolumn{2}{c}{Current system} &
    \multicolumn{2}{c}{Reform (75\%)} &
    \multicolumn{2}{c}{Reform (65\%)}\\
    \cmidrule(lr){2-3}\cmidrule(lr){4-5}\cmidrule(lr){6-7}
                                  & Rule weight & \# Rules (inc. dep.) & Rule weight & \# Rules (inc. dep.) & Rule weight & \# Rules (inc. dep.) \\
      \midrule
        Children$^\dagger$        & 30          &  3 (2)               & 11          &  2 (1)               &  11         &        2 (1)         \\
        Healthcare                & 10          &  1 (1)               &  0          &  0 (0)               &   0         &        0 (0)         \\
        Rental support            & 10          &  1 (1)               & 10          &  1 (1)               &  10         &        1 (1)         \\
        Self-employed             &  4          &  1 (0)               &  0          &  0 (0)               &   4         &        1 (1)         \\
        Labor contract            &  8          &  2 (1)               &  0          &  0 (0)               &   0         &        0 (0)         \\
        Single households         &  6          &  2 (2)               &  2          &  1 (0)               &   0         &        0 (0)         \\
        Double earner             &  4          &  1 (1)               &  4          &  1 (1)               &   4         &        1 (1)         \\
        Income brackets           &  3          &  1 (1)               &  2          &  1 (1)               &   3         &        1 (1)         \\
        Elderly                   &  6          &  1 (1)               &  0          &  0 (0)               &   4         &        1 (1)         \\
        Home value           & 6            & 1 (0)               &  0          &  0 (0)               &   0         &        0 (0)         \\
        Young handicapped         &  2          &  1 (0)               &  2          &  1 (0)               &   2         &        1 (0)         \\
        Universal benefit         &  0          &  0 (0)               &  1          &  1 (0)               &   1         &        1 (0)         \\
        \midrule
        \textbf{Total system}     & 89          & 15 (10)              & 32          &  8 (4)               &  39         &        9 (6)         \\
    \bottomrule
    \multicolumn{7}{l}{\scriptsize{$\dagger$ A full description of tax rules adopted in our system is provided in the Supplementary Material.}}
    \end{tabular}
    }
    \caption{\scriptsize {Overview of the rules included in our real-world fiscal system and the rules still active after reform. We excluded the rules pertaining to mortgages as there were no public data available. The `Rule weight' column indicates the complexity weight of all rules within a specific topic. The `Rules' column indicates the number of rules with the number of income-dependent rules between brackets. Universal rules are weighted at 1 per rate (so a universal benefit counts as 1 and a three-bracket rule as 3), group-specific rules are doubled, one unit is further added for each additional eligibility condition, and complex legacy rules carry a fixed weight of 10 (Section S1.5). Rules that serve multiple groups (Arbeidskorting; Algemene heffingskorting; see the Supplementary Material) are attributed to a single topic here, so that the rows and the total sum exactly to the 15 rules and weight of 89 reported in the Supplementary Material.}}
    \label{table: rule_topics}
\end{sidewaystable}

Policymakers have consistently expressed a desire to reform the system along three main goals: (i) eliminate `excessive' marginal pressure, (ii) design a system that is simpler and more predictable for taxpayers, and all the while (iii) protecting households from negative income shocks~\citep{Rijksoverheid2024Hervormingsagenda}. However, efforts to realize reform have been unsuccessful thus far in part due to the inability of the state for formulate promising reform candidates~\citep{eerstekamer2023over}. This section shows how such candidates could be generated using the synthetic household data and an illustrative set of constraints.

\paragraph{Generating realistic reform candidates} To generate reform candidates that target the issue of excessive marginal pressure, we start by setting income constraints that ensure no household will face a decreased net income greater than 5\% as a consequence of the reform. We can then move to the stated goals of limiting marginal pressure and reducing complexity. The former can simply be encoded as direct marginal constraints on the rates in the system. For the latter we want to balance losses in tax revenue and the complexity of the system. Thus, we make use of an iterative multi-objective setup. The first objective is to minimize tax losses (with a hard constraints in changes in total revenue greater than 1.5\%). We then add a secondary objective that includes a weighted count of the number of active rules in the system.\footnote{For this illustrative example, we use the following coding scheme to weigh rules: universal rules are counted with a weight of 1 per rate, so a universal benefit is counted as 1 and a three bracket progressive rule is counted as 3. The weight is doubled if the rule only applies to a specific group, and one unit is added for each further eligibility condition. We also allow the solver to make use of existing rules, with a fixed weight of 10. Our weighting of rules is meant to be illustrative and in practice rule weights could be determined by more precise heuristics, like perceived complexity by taxpayers or in the logistical implementation of the rule.} After solving for the first objective, we allow a further 1\% deviation in total tax revenue to further minimize this second heuristic. The result is a system that statisfies hard constraints of limiting marginal pressure and protecting household income, whilst simultaneously optimizing for a balance of minimizing tax revenue losses and minimizing the overall complexity of the system.

\begin{figure}[!t]
    \centering
    \includegraphics[width=\textwidth]{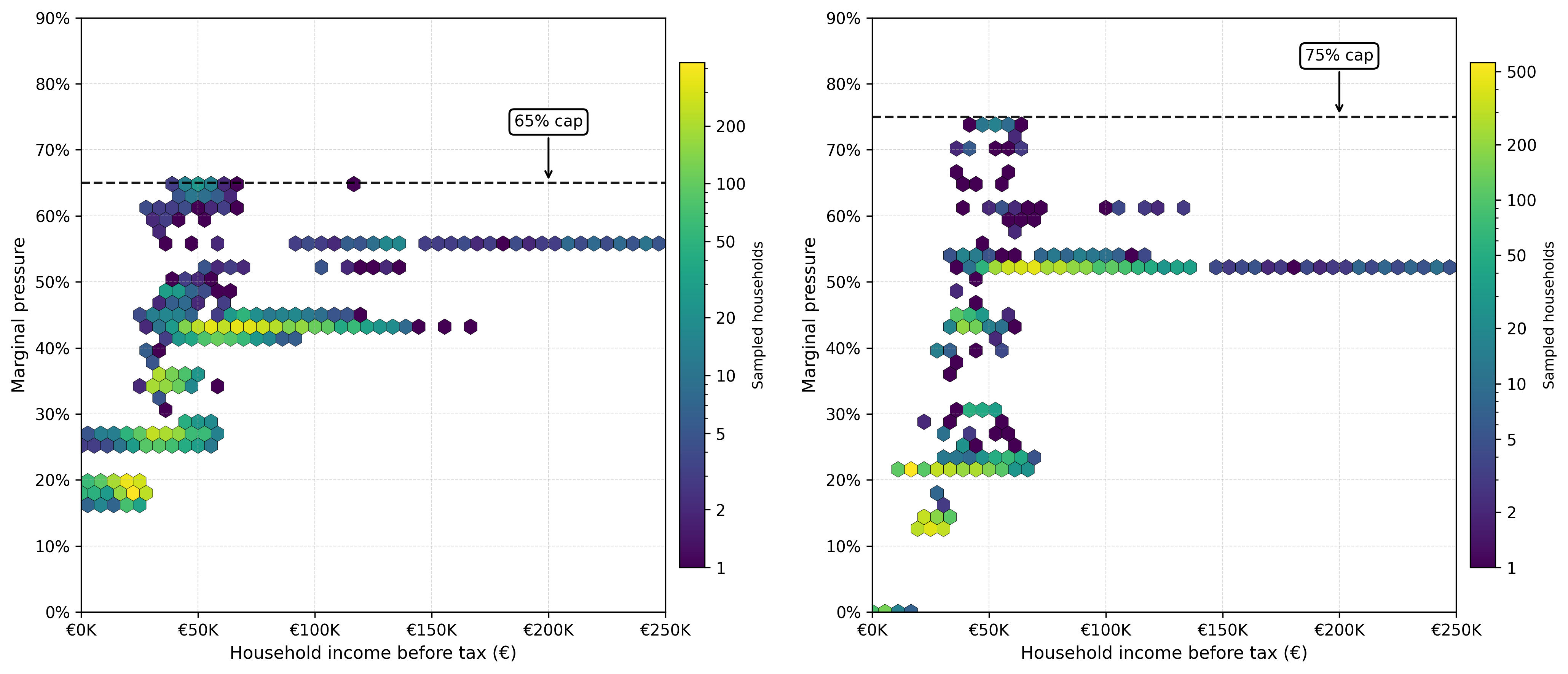}
    \caption{Illustration of two reforms with different marginal pressure constraints. A reform with a 65\% cap is given on the left and a reform with a 75\% cap is given on the right.}
    \label{fig: case_4_reform}
\end{figure}

Our approach is able to generate multiple systems that satisfy the above contraints. This is illustrated in Figure \ref{fig: case_4_intro_comp} on the right, which shows the costs and our complexity heuristic for six reforms candidates. Considerable reductions in both complexity and marginal pressure are feasible within the set constraint, although limiting marginal pressure below 65\% leads to increases in complexity to satisfy the set income constraints. Two reforms, capped at marginal pressure of 65\% and 75\%, are illustrated in Figure \ref{fig: case_4_reform}, showing considerably fewer spikes in marginal pressure. The active rules for the two systems are indicated in Table \ref{table: rule_topics}. Both reforms activate a universal benefit absent from the current code ($0\to 1$ in Table~\ref{table: rule_topics}), replacing several targeted rules with a single transfer. Because the decision variables are the legislatable parameters of the code, each certified reform \emph{is} a complete statute. The Supplementary Material prints both reformed schedules, including every bracket rate, cutoff, benefit level, and retained legacy rule. Note that for the Dutch runs the marginal guarantee is implemented as a composite-marginal-pressure cap as household income and personal income co-move (described in Equation~(S6) of the Supplementary Material). This includes the rental-support taper: rental support is retained as a legacy rule in both headline reforms (Table 4 of the Supplementary Material) and continues to withdraw with household income, so its contribution to $m_i$ is already reflected in the caps plotted in Figures~\ref{fig: case_4_intro_comp} and~\ref{fig: case_4_reform}. The figures plot the household maximum of $m_i$ across members against household income, which is possible as household-income brackets are not part of the reform support.

\begin{figure}[!t]
    \centering
    \includegraphics[width=\textwidth]{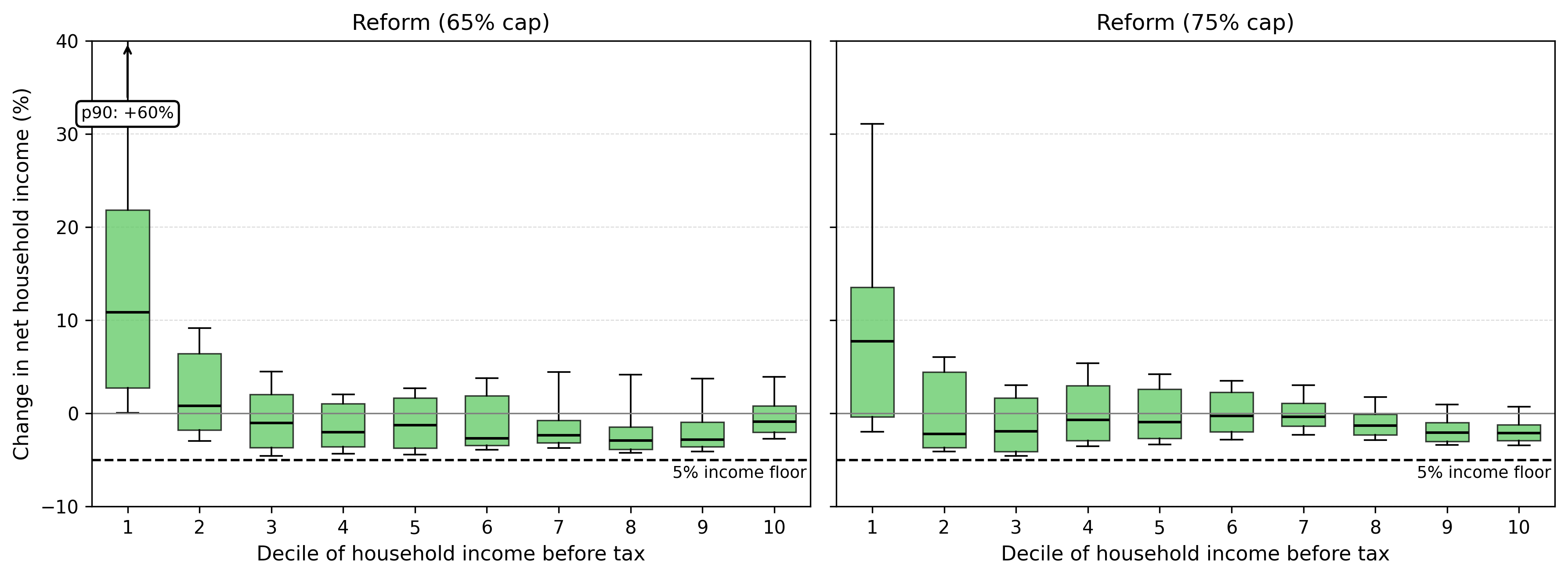}
    \caption{Distributional incidence of the two headline reforms: change in net household income (including household-level benefits) relative to the status quo, by decile of household income before tax. Boxes span the interquartile range, whiskers the 10th--90th percentiles; the dashed line marks the 5\% household income floor, which binds exactly at the largest realized losses. The 90th percentile of the bottom decile under the 65\% cap (+60\%) extends beyond the plotted range; proportional changes are largest there because status-quo net incomes are smallest. The 20 households with non-positive status-quo net income are omitted.}
    \label{fig: case_nl_incidence}
\end{figure}

Figure~\ref{fig: case_nl_incidence} shows the distributional incidence of the two reforms: the change in net household income relative to the status quo, by decile of household income before tax. The Figure directly shows the fiscal guarantees in place, with no household in any decile losing more than 5\% of net income. The aggregate cost of the reform stays inside the budget band, so the roughly 60\% of households that lose do so by small amounts that finance the large gains concentrated in the bottom decile. The applied reforms impose only the lower bound of the two-sided income constraint~\eqref{eq: income_constraint}, so individual windfalls are disciplined in aggregate through the budget band rather than capped per household. The visible upper tails in the bottom deciles are the direct consequence of this choice, and adding a per-household upper bound could further compress them if politically desirable.\footnote{The 20 households with non-positive status-quo net income are zero-gross-income, welfare-flagged single records whose recorded net income is exactly zero (16 cases) or negative by at most \euro27 (4 cases); they are retained in the optimization under the same floor and omitted only from Figure~\ref{fig: case_nl_incidence}, where proportional changes lack a meaningful base.}

\paragraph{Computational tractability} Whereas it would usually take weeks to design, model, test, and tweak a single reform candidate, this approach is able to generate candidates quickly. The Dutch instance shows that the framework is not merely expressive but computationally practical at national scale. Its decision variables are the \emph{legislatable parameters} of the code, on the order of a few dozen bracket rates over the chosen support. Importantly, they are not the taxpayers, who enter only through the constraints and the revenue objective. A formulation's size is therefore set by how rich a support and how many tax groups the analyst admits, not by how many taxpayers populate the data. The reform is shaped by the small set of constraints that bind at optimality: e.g.\ households sitting at the floor or earners at the cap. That said, further enlarging the population leaves the decision-variable count unchanged and grows only the constraint count and the per-taxpayer bookkeeping. In administrative microdata, households with identical bracketed profiles collapse to one weighted row, so the effective row count is the number of distinct profiles (Supplementary Material). Every reform reported above, including the two-stage lexicographic procedure, solves on a standard workstation to proven optimality (gap $0$). Five of the six caps solve in minutes; the $65\%$ cap takes $38$ minutes, as its cardinality stage needs longer than the nominal $30$-minute budget applied elsewhere to close (see Supplementary Material). We also report a scaling experiment that grows the sample to $1.35$ million records.

\begin{figure}[!t]
    \centering
    \includegraphics[width=\textwidth]{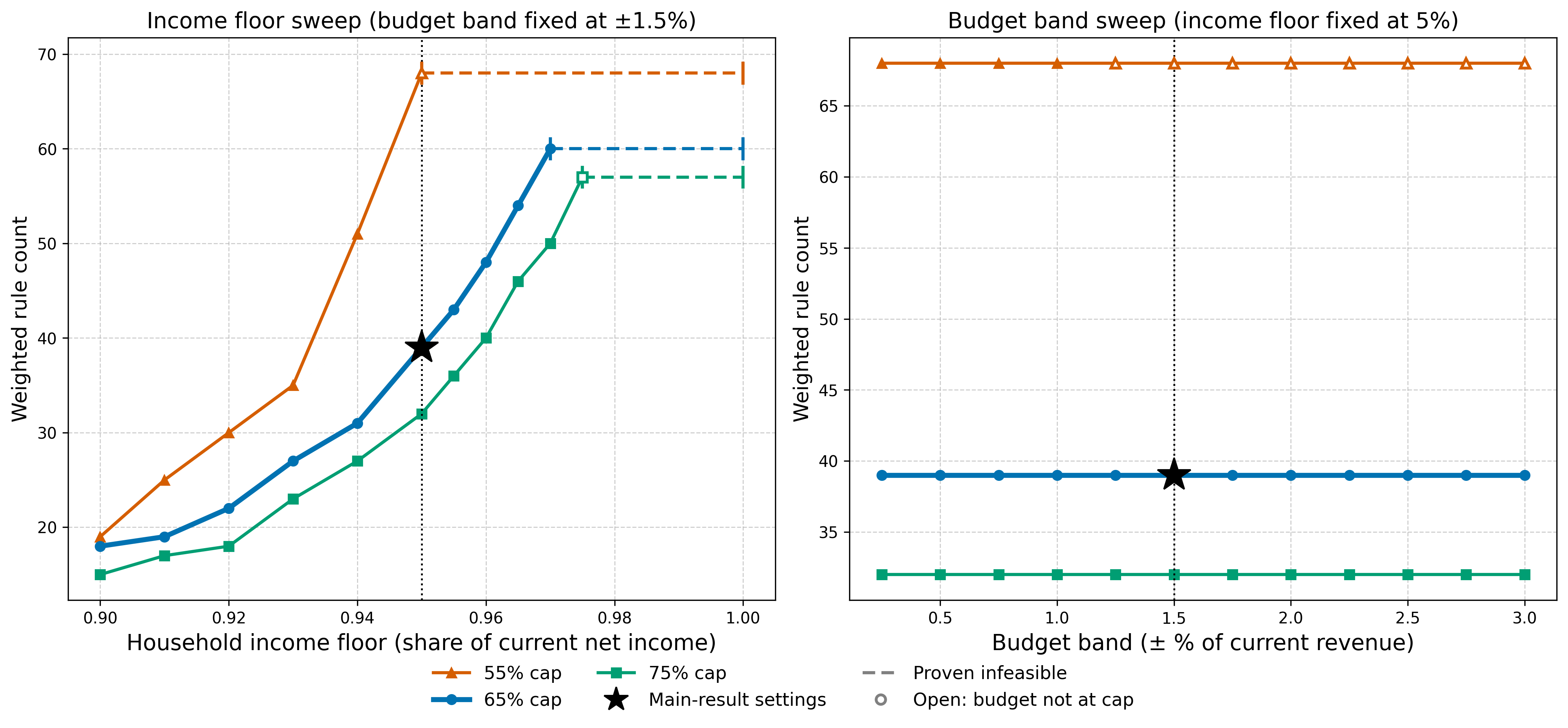}
    \caption{Complexity frontiers of the Dutch reform under marginal-pressure caps of $55\%$, $65\%$, and $75\%$. Left: weighted rule count of the cheapest conforming reform against the household income floor, with the budget band fixed at $\pm1.5\%$; dashed whiskers mark floors for which the solver proves that no conforming reform exists. Right: the same outcome against the budget band, with the income floor fixed at $5\%$. The star marks the reform of this section ($65\%$ cap, $5\%$ floor, $\pm1.5\%$ band). Filled markers indicate that the minimum revenue loss attains the revenue bound of the budget band, while open markers mark the exceptions, where the budget band is not at cap because the marginal-pressure cap itself limits how much revenue the system can raise. Tighter guarantees are thus priced in complexity, not revenue.}
    \label{fig: frontier}
\end{figure}

\paragraph{Robustness to the choice of guarantees} The reform above fixes the marginal-pressure cap, the household income floor, and the budget band at representative values, so it is natural to ask how sensitive the achievable reform is to those choices. Sweeping the income floor over $[0.90, 1.00]$ and the budget band over $[\pm0.25\%, \pm3\%]$ at caps of $55\%$, $65\%$, and $75\%$ traces the frontiers in Figure~\ref{fig: frontier}. Every combination was solved to optimality or infeasibility: of the 92 grid points, 71 were solved to optimality and 21 were certified infeasible, in 163 solver calls totaling roughly 6.4 hours of wall-clock time (per-instance statistics in the Supplementary Material). The sweep overturns the intuition that stronger guarantees are paid for in revenue. At every feasible point the cheapest conforming reform attains the revenue bound of the budget band, and widening the band does not remove a single rule (right panel). The one exception is the $55\%$ cap, where revenue capacity itself binds: beyond a $\pm1\%$ band the system cannot reach the band edge, so further budget room is left unused. Guarantees are instead paid for in complexity. The weighted rule count rises steeply as the floor tightens and rises everywhere as the cap tightens, until infeasibility is reached just above $0.95$ under a $55\%$ cap, $0.97$ under $65\%$, and $0.975$ under $75\%$. As an illustration of the diagnosis promised in Section~\ref{sec:iis}, at the infeasible point ($55\%$ cap, $0.96$ floor) the irreducible infeasible subsystem returned by the solver names 22 household income floors, the composite marginal-pressure cap, and the revenue-loss bound of the $\pm1.5\%$ budget band: under a $55\%$ cap the code cannot deliver the redistribution needed to hold those households at $96\%$ of current net income within that band, and removing any one of the three commitments restores feasibility.

\section{TaxSolver in practice}
\label{sec:practice}

Section~\ref{sec:model}'s representation and the formulations of Section~\ref{sec:reform} are realized in \texttt{TaxSolver}, an open-source, solver-agnostic Python package. All examples and results in this paper were produced with it. The package is built so that a policy analyst, rather than an optimization specialist, can assemble a statutory code from its rules, declare the fiscal guarantees and objective a reform must satisfy, and obtain either a certified reform or a certificate that none exists. The package constructs the model through a single interface and dispatches it to a backend solver: \texttt{Gurobi}~\citep{gurobi} by default, with open source alternatives supported. Below we describe the decision-support workflow the tool induces, its development and use at the Dutch Ministry of Finance, and the reproducibility package that accompanies the paper.

\paragraph{A decision-support workflow}
Casting reform as optimization reshapes the reform process itself. In the conventional sequence, a designated policy team drafts a concrete proposal that is then negotiated with stakeholders and in the broader political arena. Navigating an entire statutory code by hand means the initial proposal cannot be certified optimal against its own stated goals, so better alternatives may go undiscovered. Each objection during the political process also sends the team back to redesign and re-cost a fresh variant. The process is slow and resource-intensive, and carries no guarantee that any final proposal is adopted.

\texttt{TaxSolver} inverts this order into a workflow with three stages. The first is \emph{elicitation}: negotiators agree not on a proposal but on the guarantees and objective a reform must meet (income floors, marginal-rate caps, a budget band, protected rules), which the framework encodes directly as the constraints of Section~\ref{sec:guarantees}. The second is \emph{generation}: within those guarantees the solver returns a reform that is optimal for the stated objective, or a family of reforms as the guarantees are swept, each carrying a certificate of optimality within the modeled space rather than the analyst's assurance that no better proposal was overlooked. The third is \emph{negotiation support}. When the guarantees conflict and no reform exists, the irreducible infeasible subsystem of Section~\ref{sec:iis} names the minimal set of commitments that cannot hold at once, and re-solving under a concession shows exactly what it buys. When a concrete proposal is on the table, it can be revised in minutes rather than weeks by adjusting a guarantee, changing the objective, or fixing rules outside the reform as in Section~\ref{sec:supportchoice}. The analyst's effort thus shifts from constructing and defending individual proposals to specifying what a reform must achieve and reading off the trade-offs the optimization exposes.

\paragraph{Development and use at the Dutch Ministry of Finance}
The \texttt{TaxSolver} methodology was developed inside the Dutch Ministry of Finance, by authors working in and with the institution the method targets. This embedded development is designed so that the guarantee library of Section~\ref{sec:guarantees} encodes the commitments ministry analysts actually face in practice. The Ministry currently uses the methodology to assess reforms of the Dutch income tax code, as has been reported publicly~\citep{NRC2025, FT2025}.

\paragraph{Reproducibility}
All results in this paper are reproducible from raw data and code alone. The paper-reproduction folder of the repository contains the simulated datasets, the notebooks that generate every figure and table, and the solved-instance outputs. This includes the instrumented run scripts and per-solve records behind the computational tables of the Supplementary Material. A tagged release is permanently archived on Zenodo~\citep{taxsolver_zenodo} at \url{https://doi.org/10.5281/zenodo.21400248}, with the development version at \url{https://github.com/TaxSolver/TaxSolver}.

\section{Conclusions and future research}
\label{sec:discussion}

This paper framed tax reform as an operations-research problem: any statutory income tax code satisfying four mild assumptions reduces to a piecewise-linear function in the parameters legislation specifies (Proposition~\ref{prop:representation}), over which recovering the status quo is an inverse-feasibility problem (Proposition~\ref{prop:recovery}) and reform is a linear or mixed-integer program returning either a reform that is provably optimal within the modeled space or an infeasibility certificate that names the conflicting commitments. We demonstrated it at national scale on the Dutch income tax code, generating certified reforms that smooth marginal-rate spikes, hold household income losses within a guaranteed floor, and roughly halve the number of active rules; sweeping the guarantees shows their cost is paid in code complexity rather than revenue.

Beyond generating reforms, the framework reframes the reform process as the decision-support workflow of Section~\ref{sec:practice}: agreeing on guarantees and objectives first, generating certified alternatives rather than hand-built proposals, and using infeasibility certificates to localize political deadlocks. That the methodology was developed inside the Dutch Ministry of Finance and is in active use there to assess reforms of the income tax code (Section~\ref{sec:practice}; \citealp{NRC2025, FT2025}) is evidence that the approach is workable in the institutional setting it targets, and not only in simulation.

Several limitations bound these claims and frame the work that remains. The framework requires representative microdata covering every tax-relevant input, taxpayer characteristic, and observed liability. Settings where these data are lacking will require simulation efforts to reproduce similar data. The complexity objective relies on activation constraints (big-$M$, or solver indicator constraints in our implementation). The behavioral extension is static and reduced-form: a linearized constant-elasticity response under a bracket-stability assumption (taxpayers are presumed not to cross bracket cutoffs). It is not a structural labor-supply model. Within that model the direct nonconvex MIQCP solve certifies global optimality, although the LP-only fallback of Algorithm~\ref{alg:behavioral} is a fixed-point heuristic that carries no such certificate. Finally, our Dutch instance reproduces a real code's complexity but is calibrated only to univariate margins of the population, so its reforms illustrate the method rather than constitute concrete proposals for the Netherlands.

These limitations map onto a concrete operations-research agenda. The most immediate concerns robustness. Because each reform is computed against a single estimated microdata sample, robust optimization offers a principled way to incorporate uncertainty in the data that define its constraints~\citep{bental2009robust, gorissen2015practical}, so that the guarantees hold across a set of plausible populations rather than one realization. Beyond robustness, the dynamic-bracketing formulation invites exact breakpoint optimization in place of a fixed candidate grid. Multi-year and sequential reforms call for decomposition methods that exploit the stagewise structure noted in our multi-step discussion. Finally, replacing the reduced-form elasticity with a structural labor-supply model would ground behavioral responses in primitives and we regard the integration of such structural models~\citep{colombino2022combining} as a natural bridge between the implementation framework developed here and the optimal-tax theory it is built to serve.

\appendix
\section{Proofs}
\label{sec: proofs}

\begin{proof}[Proof of Proposition~\ref{prop:representation}]
Fix a tax group $k$. By Assumption~\ref{ass:scalar} each rule acts on a single input, so the rules partition into the incidence sets $\mathcal{R}_j$, and by Assumption~\ref{ass:additivity} the liability of taxpayer $i$ is $f_k(\mathbf{x}_i) = \sum_j \sum_{r \in \mathcal{R}_j} f_r(x_{ij};\Phi_{r,k},\alpha_{r,k}) + Z_k$. Restricted to input $j$, every $f_r$ with $r \in \mathcal{R}_j$ is continuous and piecewise linear with breakpoints $\Phi_{r,k}$ by Assumption~\ref{ass:pwl}. A sum of continuous piecewise-linear functions of the same variable is continuous and piecewise linear, with breakpoint set the union of the summands' breakpoints: on any interval containing none of those breakpoints all summands are affine, hence so is the sum. For input $j$ this gives breakpoints $\Phi_{j,k} = \bigcup_{r \in \mathcal{R}_j} \Phi_{r,k}$. Writing the resulting univariate function in the bracket basis of Definition~\ref{def:bracketing} yields $\sum_{b \in \mathcal{B}_{j,k}} \alpha_{j,b,k}\, g_b(x_{ij};\Phi_{j,k})$, where $\alpha_{j,b,k}$ is the common slope on bracket $b$. Summing over inputs and adding the group constant $Z_k$ gives~\eqref{eq: representation}. Assumption~\ref{ass:hetero} guarantees that this construction is carried out within each of the finitely many groups.
\end{proof}

\begin{proof}[Proof of Proposition~\ref{prop:recovery}]
By Proposition~\ref{prop:representation}, with the support fixed each taxpayer's liability is linear in the parameter vector $\boldsymbol{\theta}$, which stacks the bracket rates $\alpha_{j,b,k}$ and the group lump sums $Z_k$. Collecting one equality constraint $f_k(\mathbf{x}_i) = y_i'$ per taxpayer gives the linear system $\mathbf{y}' = X\boldsymbol{\theta}$, where row $i$ of $X \in \mathbb{R}^{N\times P}$ holds the bracketed inputs $g_b(x_{ij};\Phi_{j,k})$ and the group-indicator entries multiplying $\boldsymbol{\theta}$, and $P$ is the number of free parameters. Because the status quo is generated by the model the system is consistent, so its solution set is the affine subspace $\boldsymbol{\theta}^\star + \ker X$, with $\boldsymbol{\theta}^\star$ the true parameter. The solution is therefore unique, and equals $\boldsymbol{\theta}^\star$, if and only if $\ker X = \{0\}$, i.e.\ $X$ has full column rank; otherwise every element of $\boldsymbol{\theta}^\star + \ker X$ satisfies all equalities and is observationally equivalent. Two degeneracies populate $\ker X$. If, within a group, every taxpayer's input exceeds $\phi_b$, then $g_b(x_{ij};\Phi_{j,k}) = \phi_b - \phi_{b-1}$ for all such taxpayers: that column is constant and collinear with the group's lump-sum indicator column, so $\ker X$ contains a direction trading $\alpha_{j,b,k}$ off against $Z_k$. If no taxpayer's input reaches $\phi_{b-1}$, the column $g_b(\cdot;\Phi_{j,k})$ is identically zero and the corresponding unit vector lies in $\ker X$. In both cases full column rank fails precisely when a bracket-by-group cell lacks interior variation.
\end{proof}

\singlespacing
\bibliographystyle{elsarticle-harv}

\bibliography{references}

\clearpage
\setcounter{section}{0}
\setcounter{figure}{0}
\setcounter{table}{0}
\setcounter{equation}{0}
\setcounter{page}{1}   

\renewcommand{\thesection}{S\arabic{section}}
\renewcommand{\thefigure}{S\arabic{figure}}
\renewcommand{\thetable}{S\arabic{table}}
\renewcommand{\theequation}{S\arabic{equation}}

\section{Supplementary Material}

\noindent This document accompanies the manuscript ``Tax reform as a constrained optimization problem: a piecewise-linear framework and software implementation,''. Section~\ref{supp:defs} gives formal problem definitions for the illustrative examples and the Dutch case study of the main text. Section~\ref{supp:computational} reports computational details (model dimensions, solve times, and optimality gaps) for every formulation in the paper. Numbers given as plain text (e.g.\ ``Section 4,'' ``Figure 6,'' ``Eq.~(4),'' ``Proposition 1,'' ``Algorithm 1'') refer to the main text of the manuscript.

\section{Formal Problem Definitions}
\label{supp:defs}

Below we provide formal problem definitions for each of the use cases presented in Sections 4 and 5 of the main text. Throughout, we use the following notation:
\begin{itemize}
    \item $x_i$: vector of tax-relevant inputs for individual $i$, where $x_{i, \text{income}}$ indicates individual $i$'s income before tax.
    \item $y'_i$: current total tax paid by individual $i$.
    \item $\Phi$: support of the system.
    \item $\mathcal{A}$: set of decision variables (rates) to be optimized, consisting of lump sum transfers denoted by $Z$ and marginal tax rates denoted by $\alpha$.
    \item $f(x_i, \mathcal{A})|_\Phi$: total tax function evaluated at the support $\Phi$, consisting of the inner product of all inputs $x_i$ bracketed according to $\Phi$ and marginal rates vector $\alpha$, and additional tax group specific lump sum transfers $Z$ (see Eq.~(4) of the main text).
    \item $f(x_h, \mathcal{A})|_\Phi := \sum_{i \in h} f(x_i, \mathcal{A})|_\Phi$ plus household-level rules evaluated once at household inputs---the household's total tax function; $y'_h := \sum_{i \in h} y'_i$.
\end{itemize}

\subsection{Single Tax Group, Single Tax Rule}
\noindent \textbf{Tax groups:} None \\
\textbf{Inputs:} Income before tax \\
\textbf{Rules:} Income tax brackets

\paragraph{Recovery}
We solve the pure feasibility problem, requiring each individual's tax pressure to equal their current pressure:
\begin{equation}
\begin{aligned}
    \min_{\mathcal{A}}& \quad 0 \\
    \text{s.t.} \quad & f(x_i, \mathcal{A})|_\Phi = y'_i \quad \forall i
\end{aligned}
\label{eq:a_example_1_recover}
\end{equation}
where $\Phi_{\text{income}} = [25\,000, 50\,000, 75\,000, 100\,000]$. Results are shown in Figure 5a of the main text.

\paragraph{Reform}
We implement loose income constraints ensuring a $\ge 5\%$ net income increase for those earning $< 70\,000$\euro, while allowing up to a 10\% decrease for others:

\begin{equation}
\begin{aligned}
    \min_{\mathcal{A}}& \quad \sum_i [y'_i - f(x_i, \mathcal{A})] \\
    \text{s.t.} \quad & x_{i, \text{income}} - f(x_i, \mathcal{A})|_\Phi \geq (x_{i, \text{income}} - y'_i) \cdot 1.05 \quad \forall i \in [C_{i, < 70k}] \\
    & x_{i, \text{income}} - f(x_i, \mathcal{A})|_\Phi \geq (x_{i, \text{income}} - y'_i) \cdot 0.90 \quad \forall i \in [C_{i, \ge 70k}]
\end{aligned}
\label{eq:a_example_1_reform}
\end{equation}
Results are shown in Figure 5b of the main text.

\subsection{Single Tax Group, Multiple Tax Rules}
\noindent \textbf{Tax groups:} None \\
\textbf{Inputs:} Income before tax, Number of children \\
\textbf{Rules:} Income tax brackets, Healthcare benefit, Child benefit

\paragraph{Recovery} To recover the rates of the current system we start with the problem definition in Eq.~\eqref{eq:a_example_1_recover} but expand the input space with $x_{i, \text{children}}$ setting $\Phi_{\text{children}} = [0, \cdots, N_{\text{children}}]$ as its support, and expand the support of income before tax to include $[30\,000, 40\,000]$. Results are shown in Figure 6a of the main text.

\paragraph{Reform} For the second reform, we set the exact same optimization problem as Eq.~\eqref{eq:a_example_1_reform} but include that all rates are capped at 60\%:\\

\begin{equation}
\begin{aligned}
    \min_{\mathcal{A}}& \quad \sum_i [y'_i - f(x_i, \mathcal{A})] \\
    \text{s.t.} \quad & x_{i, \text{income}} - f(x_i, \mathcal{A})|_\Phi \geq (x_{i, \text{income}} - y'_i) \cdot 1.05 \quad \forall i \in [C_{i, < 70k}] \\
    & x_{i, \text{income}} - f(x_i, \mathcal{A})|_\Phi \geq (x_{i, \text{income}} - y'_i) \cdot 0.90 \quad \forall i \in [C_{i, \ge 70k}] \\
    & \alpha \leq 0.6 \quad \forall \alpha \in \mathcal{A}. \\
\end{aligned}
\label{eq:a_example_2_reform}
\end{equation}
We also exclude the rates for $x_{i, \text{children}}$ from the optimization, fixing $\alpha_{\text{children}} = 800$, and set the support of income before tax to $\Phi_{\text{income before tax}} =
[25{,}000,\allowbreak 50{,}000,\allowbreak 75{,}000,\allowbreak 100{,}000]$. Results are shown in Figure 6b of the main text.

\subsection{Multiple Tax Groups, Multiple Tax Rules}
\noindent \textbf{Tax groups:} Single / Fiscal partner, Employed / Self-employed \\
\textbf{Inputs:} Income before tax, Household income before tax, Number of children \\
\textbf{Rules:} Income tax brackets, Healthcare benefit, Child benefit, Self-employment tax credit

\paragraph{Recovery} To recover the rates of the current system we expand the problem definition in Eq.~\eqref{eq:a_example_1_recover} with the inclusion of household income before tax, with single-specific household brackets on $[30\,000, 40\,000]$ and couple-specific on $[60\,000, 75\,000]$, an additional self-employed-specific bracket on income before tax over $[0, 15\,000]$, and group-specific lump sums. The universal income schedule is shared across groups. As in the main text, the returned schedule coincides with the generating parameters---the recovery design matrix has full column rank---and we omit its figure, proceeding directly to the reform.

\paragraph{Reform} For the third reform, we set the exact same optimization problem as Eq.~\eqref{eq:a_example_1_reform} but constrain income at the level of the household and remove the group-specific rates and implement a 5\% increase in net income for households with a combined income of $85\,000$. We also remove the brackets for the healthcare benefit. This leads to the following optimization problem:

\begin{equation}
\begin{aligned}
    \min_{\mathcal{A}}& \sum_i [y'_i - f(x_i, \mathcal{A})] \\
    \text{s.t.} \quad & x_{h, \text{income}} - f(x_h, \mathcal{A})|_\Phi \geq (x_{h, \text{income}} - y'_h) \cdot 1.05 \quad \forall h \in [C_{h, < 85k}] \\
    & x_{h, \text{income}} - f(x_h, \mathcal{A})|_\Phi \geq (x_{h, \text{income}} - y'_h) \cdot 0.9 \quad \forall h \in [C_{h, \ge 85k}] \\
    & \alpha \leq 0.6 \quad \forall \alpha \in \mathcal{A}.
\end{aligned}
\label{eq:a_example_3_reform}
\end{equation}
Results are shown in Figure 7 of the main text.

\subsection{Behavioral Effects}
\noindent \textbf{Tax groups:} None \\
\textbf{Inputs:} Income before tax \\
\textbf{Rules:} Income tax brackets

We solve the same problem as Eq.~\eqref{eq:a_example_1_reform} but include behavioral effects. The objective is to minimize revenue loss subject to the income guarantees:

\begin{equation}
\begin{aligned}
    \min_{\mathcal{A}}& \quad \sum_i \bigl[y'_i - f(x^{\text{new}}_i, \mathcal{A})|_\Phi\bigr] \\
    \text{s.t.} \quad & x^{\text{new}}_{i, \text{income}} - f(x^{\text{new}}_i, \mathcal{A})|_\Phi \geq (x_{i, \text{income}} - y'_i) \cdot 1.05 \quad \forall i \in [C_{i, < 70k}] \\
    & x^{\text{new}}_{i, \text{income}} - f(x^{\text{new}}_i, \mathcal{A})|_\Phi \geq (x_{i, \text{income}} - y'_i) \cdot 0.90 \quad \forall i \in [C_{i, \ge 70k}] \\
    & x^{\text{new}}_{i, \text{income}} = x_{i, \text{income}} \cdot \left(1 + \delta \cdot \frac{\tau^{\text{sq}}_i - \tau^{\text{new}}_i}{1 - \tau^{\text{sq}}_i}\right)
\end{aligned}
\label{eq:a_example_behavioral}
\end{equation}

where $\tau^{\text{sq}}_i$ is data and reflects the active marginal rate for individual $i$ and $\tau^{\text{new}}_i$ is a decision variable and reflects the new active marginal rate for individual $i$. Note that the income floors reference the \emph{status-quo} net income $x_{i,\text{income}} - y'_i$. Under bracket stability the reformed liability at the responding income equals the liability at the original income plus the response taxed at the active marginal rate, $f(x^{\text{new}}_i, \mathcal{A})|_\Phi = f(x_i, \mathcal{A})|_\Phi + \delta\,\frac{\tau^{\text{sq}}_i - \tau^{\text{new}}_i}{1 - \tau^{\text{sq}}_i}\, x_{i,\text{income}}\, \tau^{\text{new}}_i$, which is the quadratic form the solver handles directly ($\tau^{\text{sq}}_i$ is data, so the response is linear in $\tau^{\text{new}}_i$). We sweep the elasticity over $\delta \in [0, 0.25]$. Behavioral erosion lowers the revenue collected by the cheapest conforming reform monotonically, from $7.5\%$ above status-quo revenue at $\delta = 0$ to $4.9\%$ below it at $\delta = 0.25$. Results are shown in Figure 8 of the main text.

Table~\ref{table: behavioral_convergence} validates the fixed-point fallback of Algorithm 1 of the main text against the direct solve: iterations to convergence, rate and revenue gaps, and the number of taxpayers whose responding income would cross a bracket cutoff (quantifying the bracket-stability assumption: at most 85 of 1{,}000, at the largest elasticity, predominantly moving down out of the top two brackets).

\begin{table}[!t]
  \centering
  \footnotesize
  \renewcommand{\arraystretch}{1.25}
  \begin{tabular}{r r r r r}
    \toprule
    \textbf{Elasticity $\delta$} & \textbf{Iterations} & \textbf{Rate gap (pp)} & \textbf{Revenue gap (\euro)} & \textbf{Bracket crossers} \\
    \cmidrule(lr){1-1}\cmidrule(lr){2-2}\cmidrule(lr){3-4}\cmidrule(lr){5-5}
    $0.00$ & $1$  & $0.000$ & $0$ & $0$  \\
    $0.05$ & $12$ & $0.000$ & $2{,}334$ & $20$ \\
    $0.10$ & $12$ & $0.000$ & $10{,}939$ & $46$ \\
    $0.15$ & $13$ & $0.008$ & $21{,}501$ & $59$ \\
    $0.20$ & $13$ & $0.025$ & $33{,}226$ & $72$ \\
    $0.25$ & $14$ & $0.781$ & $44{,}148$ & $85$ \\
    \bottomrule
  \end{tabular}
  \caption{Convergence of the damped fixed-point iteration of Algorithm 1 of the main text ($\lambda = 0.5$, income tolerance \euro1) across elasticities of taxable income $\delta \in [0, 0.25]$, and validation against the direct nonconvex quadratic solve, on the single-group single-rule instance with 1{,}000 taxpayers. The iteration count is the number of outer iterations (one LP solve each) to reach the income tolerance. The rate gap is the maximum absolute difference (percentage points) between the fixed-point rates and the direct-solve rates; the revenue gap is the corresponding difference in total tax revenue: the two methods bound the revenue-loss-minimizing reform slightly differently (up to $0.22\%$ of revenue at $\delta = 0.25$), while the rate schedules still nearly coincide. Bracket crossers counts taxpayers whose responding income leaves their status-quo bracket at the fixed point, quantifying the bracket-stability assumption.}
  \label{table: behavioral_convergence}
\end{table}

\subsection{The Dutch Income Tax Code Case Study}

\paragraph{Constructing the data} We rely on publicly available tax laws to identify the set of active tax rules in the Netherlands, using the most recently proposed fiscal changes for 2028 as starting point.\footnote{All tax rules in the Dutch system can be found at \url{http://www.belastingdienst.nl}. We excluded the mortgage interest deductible due to limited public data on mortgages.} The complete set of rules included in this case study are listed in Table~\ref{table: rules_nl_case}. We then simulated a set of taxpayers and households that is representative of the Dutch population on simple aggregate statistics, using publications from the Central Bureau of Statistics on population counts, incomes and household types.\footnote{\url{https://www.cbs.nl/nl-nl/nieuws/2024/25/gemiddelde-woz-waarde-woningen-3-procent-hoger}}$^,$\footnote{\url{https://longreads.cbs.nl/materiele-welvaart-in-nederland-2024/inkomen-van-huishoudens/}}$^,$\footnote{\url{https://opendata.cbs.nl/statline/\#/CBS/nl/dataset/71486ned/table?fromstatweb}} Although this dataset is not representative for the actual fiscal population in the Netherlands (discussed in more detail below) it does cover a real-world fiscal system in terms of its complexity and a true population in terms of size. Note that we use household sample weights so that our dataset contains 8,400 households and 13,500 taxpayers that can be weighted to 8.4 mln. households and 13.5 mln. taxpayers. Descriptives for the data are provided in Table~\ref{table: desc_case_4}.

\begin{table}[htbp]
\centering
  \footnotesize
\caption{Overview of Tax Rules in Dutch case}
\begin{tabular}{p{2.5cm}p{2.2cm}p{3.25cm}p{3.25cm}p{1.2cm}}
\toprule
Topic & Category & Law name & English name & Rule weight \\
& (\# brackets) & & & \\
\midrule
Children & Benefit (-)& Kinderbijslag & Child Benefit & 10$^\dagger$\\
Children & Benefit (-)& Kinderopvang-toeslag & Childcare Allowance & 10$^\dagger$\\
Children & Benefit (-)& Kindgebonden budget & Child-related Budget & 10$^\dagger$ \\
Healthcare & Benefit (-)& Zorgtoeslag & Healthcare Allowance & 10$^\dagger$ \\
Rental support & Benefit (-)& Huurtoeslag & Housing/Rental Allowance & 10$^\dagger$ \\
Home value & Deductible (3) & Eigenwoningforfait & Own home deductible & 6 \\
Elderly & Credit (3) & Ouderenkorting & Elderly Discount & 6 \\
Labor contract / Self-employed / Elderly & Credit (2) & Arbeidskorting & Labor Tax Credit & 4 \\
Self-employed & Deductible (2) & Zelfstandigenaftrek & Self-employed Deduction & 4 \\
Labor contract & Deductible (2) & Pensioenaftrek werknemer & Employee Pension Deduction & 4 \\
Single households / Double earner & Credit (2) & Algemene heffingskorting & General Tax Credit & 4 \\
Single households & Credit (2)& Inkomensafhankelijke combinatiekorting & Income-dependent Combination Credit & 4 \\
Young handicapped & Credit (0)& Jong- gehandicaptenkorting & Young Handicapped Credit & 2 \\
Single household & Credit (1) & Alleenstaande ouderenkorting & Single Elderly Discount & 2 \\
Income brackets & Brackets (3) & Belastingschijven & Tax brackets & 3\\
\bottomrule
\multicolumn{5}{l}{\footnotesize{$^\dagger$ Rule weight is set to 10 for complex rules that consist of multiple conditions.}} \\
\label{table: rules_nl_case}
\end{tabular}
\end{table}

\paragraph{Representativeness} Although our sample represents a true system's complexity, the data, and thus our reform, is not representative of the Dutch population in two important ways. First, our sample does not align with the Dutch population beyond simple univariate aggregates. For example, the overall proportion of fiscal partnerships aligns with the total population, as does the percentage of households living in a self-owned house. However, the interaction of the two is only representative if the two characteristics were completely independent from one another, which is unlikely. This limitation holds for all characteristics. In addition, we sample income distributions, asset distributions and number of children independently from other characteristics. This assumes that the distribution of e.g. the incomes of home owners is similar to those renting, which is also unlikely.

\begin{table}[!t] \centering
  \footnotesize
  \caption{Descriptive Statistics: Simulated real-world case}
  \label{table: desc_case_4}
\begin{tabular}{@{\extracolsep{5pt}}lcccccc}
\\[-1.8ex]\hline
\hline \\[-1.8ex]
Statistic & \multicolumn{1}{c}{N} & \multicolumn{1}{c}{Mean} & \multicolumn{1}{c}{St. Dev.} & \multicolumn{1}{c}{Median} & \multicolumn{1}{c}{Min} & \multicolumn{1}{c}{Max} \\
\hline \\[-1.8ex]
Income before tax & 13,500 & 32{,}861 & 25{,}584 & 27{,}460 & 0 & 312{,}396 \\
Home value & 13,500 & 209{,}886 & 191{,}959 & 314{,}148 & 0 & 558{,}545 \\
Assets & 13,500 & 57{,}781 & 58{,}459 & 37{,}903 & 0 & 380{,}230 \\
Monthly rent & 13,500 & 530 & 647 & 0 & 0 & 2{,}566 \\
Social rent  & 1,400 & 10,37\% \\
Income: benefits & $2,067$ & 15.3\% \\
Income: employment & $10,573$ & 78.3\% \\
Income: self-employed & $860$ & 6.4\% \\
Wealth: wealthy$^\dagger$ & $4,296$ & 31.8\% \\
Wealth: not wealthy$^\dagger$ & $9,204$ & 68.2\% \\
Fiscal partner: yes & $10,200$ & 75.6\% \\
Fiscal partner: no & $3,300$ & 24.4\% \\
Pension age: yes & $3,319$ & 24.6\% \\
Pension age: no & $10,181$ & 75.4\% \\
Number of children & 13,500 & 0,57 & 0,80 & 0 & 0 & 5 \\
Weight & 13,500 & 1{,}000 & 0 & 1{,}000 & 1{,}000 & 1{,}000 \\
\hline \\[-1.8ex]
\multicolumn{7}{l}{\footnotesize{$^\dagger$Wealth is defined as having assets above \euro 57\,000.}} \\
\end{tabular}
\end{table}

Second, the most important input in the fiscal system, pre-tax income, is based on publicly available income percentiles. These do not contain aggregate statistics for the top income percentile, which means we do not accurately represent the very top incomes.\footnote{The top income percentile is simply represented by ``more than X\euro'' in public statistics on personal incomes.} Overall, this case study accurately represents individual tax rules but does not represent macro outcomes, like total tax revenue.
\newline
\newline
\noindent \emph{Reform setup}. This reform differs from those before in that a support is defined that is not equal to or a subset from the current system's support. The general strategy is to provide the solver with a flexible support that allows many degrees of freedom to optimize the system, while using a heuristic for the number of active rules to reduce complexity. As a consequence, rich supports are defined for both the entire taxpayer population as well as for specific tax groups. We provide the solver with the same tax groups as encountered in the current system. The reported runs additionally impose ascending universal bracket rates: at most five active universal brackets with a minimum gap of three cutoffs, at most three active group-specific brackets with a minimum gap of one and the last forced to rate zero, and mutually exclusive rule pairs (e.g., a new versus legacy child benefit).
\newline
\newline
\noindent \textbf{Tax groups:} Single / Fiscal partner, Employed / Self-employed / Benefits, Wealthy / Not wealthy, Social renter / Renter / Homeowner, Young handicapped / Not Young handicapped, Retiree / Non-retiree \\
\textbf{Inputs:} Income before tax, Household income before tax, Number of children (any age, aged 0-5, aged 6-11, aged 12-15, and aged 16-17), Home value \\
\textbf{Rules:} All existing rules deemed `complex' and a custom support for reform (see below)

To reform this system we run the lexicographic two-stage procedure of Section 4.4 of the main text for several levels of the marginal-pressure cap $\gamma$. Let $\mathcal{F}(\gamma)$ be the feasible set defined by the fiscal guarantees,
\begin{equation}
\mathcal{F}(\gamma) = \left\{\, \mathcal{A} \;:\;
\begin{aligned}
    & x_h - f(x_h, \mathcal{A})|_\Phi \geq (x_h - y'_h)\cdot 0.95 && \forall h,\\
    & m_i(\mathcal{A}) \leq \gamma && \forall i,\\
    & 0.985\textstyle\sum_i w_i y'_i \;\leq\; \sum_i w_i f(x_i, \mathcal{A}) \;\leq\; 1.015\textstyle\sum_i w_i y'_i,\\
    & \underline{\alpha}_b \le \alpha_b \le \overline{\alpha}_b && \forall b
\end{aligned}
\,\right\},
\label{eq:a_dutch_feasible}
\end{equation}
In the reported runs, $\underline{\alpha}_b = -0.5$ and $\overline{\alpha}_b = 0.8$ for the universal schedule and $[\underline{\alpha}_b, \overline{\alpha}_b] = [-0.5, 0]$ for group-specific schedules, with $0 \le m_i \le 1$.
Here $m_i(\mathcal{A})$ is a composite marginal-pressure guarantee: for earner $i$, let $\mathcal{J}(i)$ be the inputs that co-move one-for-one with $i$'s own earnings (in this case household income and personal income). Let $b_j(x_i)$ be the bracket of input $j$ occupied at the observed $x_i$, and let $\mu_r(x_i)$ be the status-quo marginal pressure of retained legacy rule $r$ with respect to a one-euro increase in $i$'s own earnings (zero for rules keyed to inputs outside $\mathcal{J}(i)$).

The composite marginal pressure and its cap are then defined as:
\begin{equation}
m_i \;=\; \sum_{j \in \mathcal{J}(i)} \alpha^{(j)}_{b_j(x_i)}
\;+\; \sum_{r \in \mathcal{L}} s_r\,\mu_r(x_i),
\qquad m_i \le \gamma \quad \forall i,
\label{eq: composite_cap}
\end{equation}
which is linear in $(\alpha, s)$ because bracket membership is evaluated at the observed inputs (and remains so under the bracket-stability assumption of Section~4.5): the sum of the consolidated rate on every input that co-moves with taxpayer $i$'s own earnings and the $s_r$-scaled marginal pressure of every retained legacy rule, including the rental-support taper reused below. The feasible set $\mathcal{F}(\gamma)$ thereby guarantees that no household loses more than $5\%$ of net income, that no taxpayer's composite marginal pressure exceeds $\gamma$, and that total revenue stays within $\pm1.5\%$ of the status quo. The first stage minimizes revenue loss over this set,
\begin{equation}
L^\star(\gamma) = \min_{\mathcal{A}\,\in\,\mathcal{F}(\gamma)} \; \sum_i w_i\bigl[y'_i - f(x_i, \mathcal{A})\bigr].
\label{eq:a_dutch_stage1}
\end{equation}
The second stage minimizes the weighted rule count, allowing revenue loss to grow by a further slack $\beta = 0.01\sum_i y'_i$ beyond the first-stage optimum:
\begin{equation}
\begin{aligned}
    \min_{\mathcal{A},\,z,\,s}\quad & \sum_b w_b\, z_b\\
    \text{s.t.}\quad & \mathcal{A} \in \mathcal{F}(\gamma),\\
    & \sum_i w_i\bigl[y'_i - f(x_i, \mathcal{A})\bigr] \;\leq\; L^\star(\gamma) + \beta,\\
    & 0.9\, z_r \;\leq\; s_r \;\leq\; 1.1\, z_r, \quad z_r \in \{0,1\} && \forall r \in \mathcal{L},
\end{aligned}
\label{eq:a_dutch_stage2}
\end{equation}
All sums over taxpayers and households carry the sampling weights $w_i$. In the Dutch sample the weights are uniform at $0.001$ per record (each record represents $1{,}000$ taxpayers, so $\sum_i w_i = 13.5$ corresponds to $13.5$ million), and the unweighted row sums coincide with the weighted ones up to this common factor.
Here, $z_b$ activates candidate rate $\alpha_b$, the weights $w_b$ follow the weighting scheme described in Section 5 of the main text, $\mathcal{L}$ is the set of legacy rules the solver may reuse, and $s_r$ scales a retained legacy rule (its activation $z_r$ entering the objective with the legacy weight). Activation is posed as solver indicator constraints: $z_b = 0 \Rightarrow \alpha_b = \alpha_{b-1}$ (for first brackets and benefits, $\alpha_b = 0$) and $z_b = 1 \Rightarrow \underline{\alpha}_b \le \alpha_b \le \overline{\alpha}_b$; an explicit big-$M$ linearization is valid with $M$ equal to the width of the rate box. The support, tax groups, and legacy rules entering $\mathcal{F}(\gamma)$ are specified next.

We provide the following basic support for income before tax.
\[
\begin{split}
\Phi_{\text{income before tax}} =
[&0, 10\,000, 20\,000, 25\,000, 30\,000, 35\,000, 50\,000, 70\,000, 90\,000,\\
&110\,000, 150\,000, 200\,000, 250\,000, \infty]
\end{split}
\]
and provide group-specific rates for the following tax groups: the self-employed, the lowest earner in a fiscal partnership, young handicapped individuals, retirees, and those in conventional employment on the interval
\[
\Phi_{k,\text{income before tax}} =
[0, 20\,000, 30\,000, 50\,000, 90\,000, 150\,000, 250\,000].
\]
This leads to 13 universal rates and $5 \times 6 = 30$ group-specific rates for income before tax.

We then allow both personal and household level benefits for all of the tax groups mentioned above individually, as well as combinations with the `not wealthy' group, leading to 10 group-specific benefits. We then allow single brackets for all other inputs with a simple linear support.

In addition to the standard setup to populate $f(x_i, \mathcal{A})|_\Phi$, we allow the solver to directly make use of complex existing rules in the system, such as the healthcare benefit and child benefits (denoted with a weight factor of 10 in Table~\ref{table: rules_nl_case}), by including the current absolute and marginal tax pressure as fixed inputs in the total tax function with a scaling factor on $[0.9, 1.1]$ and binary activation variable. This allows the solver to either use the existing complex rules as is, scale them up or down, or omit them entirely.

The two stages instantiate the lexicographic program~\eqref{eq:a_dutch_stage1}--\eqref{eq:a_dutch_stage2}: the first finds the most cost-effective reform, and the second spends a further $1\%$ of revenue ($\beta$) to minimize the weighted rule count, with weights $w_b$ as described in Section 5 of the main text. Equations~\eqref{eq:a_dutch_feasible}--\eqref{eq:a_dutch_stage2}, instantiated with this support and group inventory, constitute the complete Dutch MILP: 77 candidate rules, $13{,}580$ continuous and $77$ binary variables, and $48{,}972$ constraint rows, itemized in Section~\ref{supp:computational}.

The above is repeated for $\gamma \in [0.55, 0.6, 0.65, 0.7, 0.75, 0.8]$. Results are shown in Figures 9 and 10 of the main text.

\paragraph{The reformed schedules in full} Table~\ref{table: reform_schedules} prints the complete certified schedules of the two headline reforms ($\gamma = 0.65$ and $\gamma = 0.75$): the universal bracket rates and cutoffs, the group-specific rates with their income ranges, the benefit levels, and the scaling factors of the retained legacy rules. Everything not listed is deactivated by the cardinality stage. The table makes concrete the claim that a solution is a set of directly legislatable numbers. Each reform is fully specified by roughly a dozen parameters, extracted verbatim from the solved rule tables in the reproducibility package.

\begin{table}[!t]
  \centering
  \footnotesize
  \resizebox{\textwidth}{!}{%
  \begin{tabular}{lrrrr}
    \toprule
    & \multicolumn{2}{c}{Reform (65\% cap)} & \multicolumn{2}{c}{Reform (75\% cap)} \\
    \cmidrule(lr){2-3}\cmidrule(lr){4-5}
    & Income range & Rate & Income range & Rate \\
    \midrule
    \multicolumn{5}{l}{\emph{Universal income tax brackets (per person, income before tax)}} \\
    \quad & \euro{}0--\euro{}30{,}000 & 27.3\% & \euro{}0--\euro{}10{,}000 & 0.0\% \\
    \quad & \euro{}30{,}000--\euro{}90{,}000 & 43.4\% & \euro{}10{,}000--\euro{}35{,}000 & 21.7\% \\
    \quad & above \euro{}90{,}000 & 55.2\% & above \euro{}35{,}000 & 52.3\% \\
    \addlinespace
    \multicolumn{5}{l}{\emph{Group-specific rates (per person, income before tax)}} \\
    \quad Lowest earner in fiscal partnership & \euro{}0--\euro{}50{,}000 & -8.2\% & \euro{}20{,}000--\euro{}50{,}000 & -8.3\% \\
    \quad Retirees (AOW) & \euro{}0--\euro{}50{,}000 & -1.6\% & --- & --- \\
    \quad Self-employed & \euro{}0--\euro{}30{,}000 & -2.1\% & --- & --- \\
    \addlinespace
    \multicolumn{5}{l}{\emph{Benefits (\euro{} per year)}} \\
    \quad Universal per-person benefit & \multicolumn{2}{c}{\euro{}4{,}534} & \multicolumn{2}{c}{\euro{}2{,}067} \\
    \quad Household benefit per child & \multicolumn{2}{c}{\euro{}8{,}962} & \multicolumn{2}{c}{\euro{}8{,}923} \\
    \quad Young-handicapped benefit & \multicolumn{2}{c}{\euro{}941} & \multicolumn{2}{c}{\euro{}928} \\
    \quad Single-household benefit & \multicolumn{2}{c}{---} & \multicolumn{2}{c}{\euro{}366} \\
    \addlinespace
    \multicolumn{5}{l}{\emph{Retained legacy rules (scaling of current statute)}} \\
    \quad Child-related budget (kindgebonden budget) & \multicolumn{2}{c}{$\times1.10$} & \multicolumn{2}{c}{$\times1.10$} \\
    \quad Rental support (huurtoeslag) & \multicolumn{2}{c}{$\times0.96$} & \multicolumn{2}{c}{$\times0.96$} \\
    \bottomrule
  \end{tabular}%
  }
  \caption{The two headline certified reforms of the Dutch case in full (65\% and 75\% marginal-pressure caps; both guarantee a 5\% household income floor within a $\pm$1.5\% budget band). Group-specific rates are additive to the universal brackets and apply on the stated income range only (a negative rate is an income-dependent credit). Retained legacy rules keep the current statutory formula, scaled by the stated factor. All remaining candidate and current rules, including the healthcare allowance, child benefit, childcare allowance, labor tax credit, general tax credit, elderly discounts, and the own-home deductible, are deactivated (Table 2 of the main text). Extracted from the solved instances underlying Figure 10 of the main text.}
  \label{table: reform_schedules}
\end{table}

\section{Computational Details}
\label{supp:computational}

All computational results in the paper were obtained with Gurobi~12.0.0 through the \texttt{TaxSolver} interface, on a single workstation (Apple M4~Max, 16 cores, 64\,GB RAM, macOS~15.7). Unless stated otherwise the solver ran with default parameters on 16 threads. The Dutch instances additionally set \texttt{NumericFocus}=2 and a nominal wall-clock limit of $1{,}800$ seconds per solve. The default relative MIP-gap tolerance of $10^{-4}$ was never loosened. Table~\ref{table: performance} reports, for every formulation in the paper, the model dimensions as passed to the solver, the wall-clock solve time, and the terminal optimality gap. Table~\ref{table: scaling} reports the scaling experiment in the number of sampled records. The instrumented run scripts and the per-solve records are part of the reproducibility package (Section 6 of the main text).

\begin{table}[!t]
  \centering
  \footnotesize
  \renewcommand{\arraystretch}{1.0}
  \setlength{\tabcolsep}{3pt}
  \begin{tabular}{l l r r r r r}
    \toprule
    \textbf{Formulation} & \textbf{Class} & \textbf{Cont.} & \textbf{Bin.} & \textbf{Constr.} & \textbf{Time (s)} & \textbf{Gap} \\
    \midrule
    \multicolumn{7}{l}{\emph{Illustrative examples} (Section 4.7 of the main text; $1{,}000$ taxpayers)} \\
    \addlinespace[2pt]
    \quad Single rule, recovery            & LP   & $7$       & $5$   & $3{,}004$  & $<0.1$ & $0$ \\
    \quad Single rule, reform              & LP   & $1{,}008$ & $5$   & $4{,}433$  & $<0.1$ & $0$ \\
    \quad Multiple rules, recovery         & LP   & $11$      & $9$   & $3{,}004$  & $<0.1$ & $0$ \\
    \quad Multiple rules, reform           & LP   & $1{,}010$ & $7$   & $4{,}433$  & $<0.1$ & $0$ \\
    \quad Multiple groups, recovery        & LP   & $1{,}014$ & $11$  & $3{,}673$  & $<0.1$ & $0$ \\
    \quad Multiple groups, reform          & LP   & $1{,}010$ & $7$   & $3{,}893$  & $<0.1$ & $0$ \\
    \quad Dynamic bracketing               & MILP & $1{,}034$ & $31$  & $4{,}460$  & $<0.1$ & $8\cdot10^{-5}$ \\
    \addlinespace[2pt]
    \multicolumn{7}{l}{\emph{Behavioral responses} (Section 4.5 of the main text; shown at $\delta = 0.25$)} \\
    \addlinespace[2pt]
    \quad Direct solve, per stage          & MIQCP & $3{,}008$ & $5$   & $4{,}008$\,{+}\,$2{,}000$q & $<0.2$ & $\le 10^{-4}$ \\
    \quad Alg.\ 1 (main text), per LP iterate & LP & $1{,}008$ & $5$ & $4{,}008$ & $<0.1$ & $0$ \\
    \quad Alg.\ 1 (main text), total run & --- & --- & --- & --- & $0.13$--$2.0$ & --- \\
    \midrule
    \multicolumn{7}{l}{\emph{Dutch income tax code} (Section 5 of the main text; $13{,}500$ taxpayers, $8{,}400$ households)} \\
    \addlinespace[2pt]
    \quad Reform $\gamma=0.55$ & MILP & $13{,}580$ & $77$ & $48{,}972$ & $42$ / $144$ & $0$ / $0$ \\
    \quad Reform $\gamma=0.60$ & MILP & $13{,}580$ & $77$ & $48{,}972$ & $53$ / $1{,}714$ & $0$ / $0$ \\
    \quad Reform $\gamma=0.65$ & MILP & $13{,}580$ & $77$ & $48{,}972$ & $59$ / $2{,}252^{\ast}$ & $0$ / $0$ \\
    \quad Reform $\gamma=0.70$ & MILP & $13{,}580$ & $77$ & $48{,}972$ & $52$ / $890$ & $0$ / $0$ \\
    \quad Reform $\gamma=0.75$ & MILP & $13{,}580$ & $77$ & $48{,}972$ & $133$ / $476$ & $0$ / $0$ \\
    \quad Reform $\gamma=0.80$ & MILP & $13{,}580$ & $77$ & $48{,}972$ & $2.6$ / $523$ & $0$ / $0$ \\
    \bottomrule
  \end{tabular}
  \caption{Problem size and solver performance for every formulation in the paper: continuous (Cont.) and binary (Bin.) variables, linear constraint rows (``{+}\,$n$q'' marks quadratic constraints), wall-clock time, and terminal relative MIP gap. Dutch rows show ``stage~1 / two-stage'' entries; every stage solves to proven optimality (gap $0$). $^{\ast}$~exceeds the nominal $1{,}800$\,s budget used elsewhere; confirmed optimal in an extended run (see main text above). Solver, parameters, and hardware as described above.}
  \label{table: performance}
\end{table}

Two reporting conventions apply throughout. First, the implementation attaches an activation indicator to every candidate rule, so even the formulations classified LP are passed to the solver with a handful of binary variables. These indicators do not enter the objective in the LP-class formulations, and all such instances solve at gap $0$ in well under a second. Second, the behavioral rows of Table~\ref{table: performance} are per elasticity: the full sweep of Figure 8 of the main text solves six elasticities (six direct MIQCP solves) and six fixed-point runs of Algorithm 1 of the main text comprising $1$--$14$ LP solves each.

\paragraph{Model dimensions of the Dutch instance} The decision variables of the Dutch reform are fixed by the support and group inventory above: 13 universal bracket rates, $5 \times 6 = 30$ group-specific bracket rates, 28 candidate benefit and household-benefit rates (the group-specific benefit families of the setup above, including their wealth-capped and single-topup variants), one imputed-rent rate, and five legacy-rule scalings. This gives 77 candidate rules in total, each carrying one continuous rate and one binary activation variable ($80$ non-taxpayer continuous columns in all). The remaining continuous columns are bookkeeping at the base sample size $N=13{,}500$: one auxiliary variable per taxpayer that is introduced to link each taxpayer's effective marginal rate (a linear combination of active rule rates) to the scalar cap $\gamma$ via Gurobi's \texttt{MAX} general constraint. Three global scalars are added for aggregate post-reform revenue, revenue change (the stage-1 objective), and \texttt{highest\_marginal\_pressure}. Together these yield $77 + 13{,}500 + 3 = 13{,}580$ continuous and $77$ binary variables.

The linear constraint count is dominated by the microdata: $8{,}400$ household net-income guarantees (the 5\% loss limit) and $3 \times 13{,}500 = 40{,}500$ taxpayer marginal-accounting rows (two bound constraints $0 \le m_i \le 1$ plus one equality linking each marginal expression to its \texttt{MAX} auxiliary). The remaining $72$ structural rows encode the formulation itself. $23$ do so on the universal income schedule (12 ascending-bracket inequalities, one cardinality cap at five active brackets, ten minimum-gap windows). $35$ do so on the five group-specific schedules (five cardinality caps, 25 minimum-gap windows, five terminal-bracket-zero conditions). Nine on mutual-exclusion rows across legacy and candidate benefit families. Four on budget-band rows (two definitional equalities plus upper and lower revenue limits at $\pm 1.5\%$) and one row capping \texttt{highest\_marginal\_pressure} at $\gamma$, for $8{,}400 + 40{,}500 + 72 = 48{,}972$ linear rows in total. Gurobi additionally reports $83$ general constraints, which lie outside the linear row count reported in Table~\ref{table: performance}. Because the income guarantees are stated per household, ``national scale'' in the main text refers to the size of the represented population ($13.5$ million taxpayers via sample weights), not the row count. The scaling experiment below addresses the row count directly.

\paragraph{Optimality gaps} The recovery and reform LPs of Section 4 of the main text solve to optimality (gap $0$). For the Dutch instances, the stage-1 revenue-loss MILP solved to proven optimality (gap $0$) at every cap $\gamma$, as did the cardinality stage of the two-stage lexicographic procedure (executed through Gurobi's sequential multi-objective mode, each stage run to the default $10^{-4}$ relative gap tolerance). Eleven of the twelve stages across the six-cap sweep closed within the nominal $1{,}800$-second budget. For the twelfth, the cardinality stage at $\gamma = 0.65$ needed a separate run with the limit relaxed to four hours (incumbent and proven bound both $39$, gap $0$, at $2{,}252$ seconds), confirming that the headline reform at that cap is a certified global optimum. The direct behavioral solves are nonconvex (bilinear) MIQCPs, which Gurobi handles with its global spatial branch-and-bound algorithm by default (\texttt{NonConvex}$=-1$). Every solve reached proven global optimality within the default tolerance, with terminal gaps below $10^{-4}$. Table~\ref{table: performance} reports the direct solve against Algorithm 1 of the main text, whose per-iteration LPs solve in milliseconds so that the full damped iteration completes in seconds.

\paragraph{Instances behind the guarantee-sweep figure} The guarantee sweep of Figure 12 of the main text comprises 92 grid points: 16 income-floor points and 12 budget-band points at each of the caps $55\%$, $65\%$, and $75\%$, plus an 8-point cap sweep. Each feasible point is solved twice (a single-objective revenue-loss solve, then the two-stage procedure), and each infeasible point once, for 163 solves in total: 71 points solved to optimality and 21 were proved infeasible. The 92 revenue-loss solves took between $0.8$ and $615$ seconds (median $9$) of which the 71 feasible ones all at gap $0$ and the 71 two-stage solves took between $26$ seconds and $60$ minutes (median $155$ seconds). Total wall-clock time for the full sweep was approximately $6.4$ hours.

\paragraph{Scaling in the number of records} A formulation's decision-variable count is set by the support and tax groups, not by the data, so enlarging the sample grows only the bookkeeping variables and the constraint count. Table~\ref{table: scaling} profiles the Dutch reform at $\gamma = 0.65$ as the sample is replicated tenfold and a hundredfold (weights divided by the replication factor, leaving all weighted aggregates unchanged): $N = 13{,}500 \rightarrow 135{,}000 \rightarrow 1{,}350{,}000$ taxpayer records. Model construction remains negligible relative to solve time at every size. The revenue-loss stage solves to proven optimality throughout, in $56$, $851$, and $12{,}099$ seconds respectively. The cardinality stage returns feasible reforms certified at growing gaps under a matched $30/30/240$-minute time budget ($12.8\%$, $38.1\%$, and $78.6\%$). The $N=13{,}500$ row matches the $\gamma=0.65$ entry of Table~\ref{table: performance} and the $12.8\%$ gap is a matched-budget comparison across $N$, not evidence that the base instance fails to reach proven optimality. At $N = 1{,}350{,}000$ the stage-1 time in Table~\ref{table: scaling} is taken from inside the two-stage run. A stand-alone stage-1 attempt on the same instance proved the same lower bound but found no incumbent within its $14{,}400$-second limit, illustrating solver variability on the highly degenerate replicated instance, whose $100$-fold row symmetry is itself adversarial for branch-and-bound. This is precisely the regime in which the aggregation observation below applies.

\begin{table}[!t]
  \centering
  \footnotesize
  \renewcommand{\arraystretch}{1.25}
  \setlength{\tabcolsep}{4pt}
  \begin{tabular}{r r r r r r r}
    \toprule
    \textbf{Records $N$} & \textbf{Rates} & \textbf{Bin.} & \textbf{Cont.} & \textbf{Constr.} & \textbf{Stage 1 (s)} & \textbf{Two-stage (s / gap)} \\
    \midrule
    $13{,}500$      & $77$ & $77$ & $13{,}580$      & $48{,}972$      & $56$        & $1{,}800$ / $0.128$ \\
    $135{,}000$     & $77$ & $77$ & $135{,}080$     & $489{,}072$     & $851$       & $1{,}800$ / $0.381$ \\
    $1{,}350{,}000$ & $77$ & $77$ & $1{,}350{,}080$ & $4{,}890{,}072$ & $12{,}099^{\dagger}$ & $14{,}758$ / $0.786$ \\
    \bottomrule
  \end{tabular}
  \caption{Scaling of the Dutch reform ($\gamma = 0.65$) as the number of sampled taxpayer records $N$ grows by two orders of magnitude (base sample replicated $10\times$ and $100\times$ with weights divided by the replication factor, so all weighted aggregates are unchanged). The decision variables ($77$ candidate rule rates and their activation binaries) are invariant in $N$; the bookkeeping variables and constraint rows grow linearly. Stage 1 (revenue loss) solved to proven optimality at every size; the two-stage column reports wall-clock seconds and the terminal gap of the cardinality stage ($^{\dagger}$see the main text above for time limits and details). Solver and hardware as in Table~\ref{table: performance}.}
  \label{table: scaling}
\end{table}

\paragraph{Aggregating identical households} A household enters the model only through its bracketed inputs, group memberships, and status-quo pressures, so households with identical profiles contribute identical constraint rows and identical terms to the revenue objective. Such duplicates can be aggregated into a single row with a summed weight, making the effective row count the number of \emph{distinct} household profiles rather than the number of records. In administrative microdata, where many households share bracketed inputs and group membership, this aggregation can compress the instance substantially. In our simulated sample all $8{,}400$ base households are distinct (incomes are drawn from continuous distributions) and the replicated instances of Table~\ref{table: scaling} would collapse back to $8{,}400$ effective rows under aggregation.

\end{document}